## Research Paper

# Turbulence measurements in the neutral ISM from HI-21 cm emission-absorption spectra

Atanu Koley[1]

[1]Departamento de Astronomía, Universidad de Concepción, Casilla 160-C, Concepción, Chile

**Abstract**

We study the correlation between the non-thermal velocity dispersion ($\sigma_{nth}$) and the length-scale ($L$) in the neutral interstellar medium (ISM) using a large number of HI gas components taken from various published HI surveys and previous HI studies. We notice that above the length-scale ($L$) of 0.40 pc, there is a power-law relationship between $\sigma_{nth}$ and $L$. However, below 0.40 pc, there is a break in the power-law, where $\sigma_{nth}$ is not significantly correlated with $L$. It has been observed from the Markov chain Monte Carlo (MCMC) method that for the dataset of $L > 0.40$ pc, the most probable values of intensity ($A$) and power-law index ($p$) are 1.14 and 0.55 respectively. Result of $p$ suggests that the power-law is steeper than the standard Kolmogorov law of turbulence. This is due to the dominance of clouds in the cold neutral medium. This is even more clear when we separate the clouds into two categories: one for $L > 0.40$ pc and the kinetic temperature ($T_k$) is $< 250$ K, which are in the cold neutral medium (CNM) and for other one where $L > 0.40$ pc and $T_k$ is between 250 K and 5000 K, which are in the thermally unstable phase (UNM). Most probable values of $A$ and $p$ are 1.14 and 0.67 respectively in the CNM phase and 1.01 and 0.52 respectively in the UNM phase. A greater number of data points is effective for the UNM phase in constructing a more accurate estimate of $A$ and $p$, since most of the clouds in the UNM phase lie below 500 K. However, from the value of $p$ in the CNM phase, it appears that there is a significant difference from the Kolmogorov scaling, which can be attributed to a shock-dominated medium.

**Keywords:** ISM: atoms, ISM: kinematics and dynamics, ISM: lines and bands, ISM: structure, Turbulence

(Received xx xx xxxx; revised xx xx xxxx; accepted xx xx xxxx)

## 1. Introduction

Thermal stability analysis shows that neutral interstellar medium (ISM) is bistable in nature. The cold neutral medium (CNM, $T_k \sim$ 40 - 250 K) is embedded in the warm neutral medium (WNM, kinetic temperature $T_k \geq 5000$ K) in rough thermal pressure ($P_{th}$) equilibrium. Gas in the intermediate phase (250 K $< T_k <$ 5000 K) is thermally unstable and drifts into these stable phases (CNM & WNM) with slight perturbation (Murray *et al.* 2018; Roy *et al.* 2013a,b; Heiles & Troland 2003a,b; Wolfire *et al.* 2003, 1995; Field *et al.* 1969; Field 1965). However, observational studies have suggested that the ISM may be multiphase in nature (Roy *et al.* 2013b; Heiles & Troland 2003b,a). Gas in the intermediate phase or thermally unstable phase (UNM) can remain stable for a long period. Additionally, it has been suggested that the amount of gas in the UNM phase is strongly correlated with the level of turbulence (Audit & Hennebelle 2005). Therefore, it is a very interesting task to study the nature of turbulence in neutral ISM in terms of its intensity and scale-dependent power-law. Homogeneous, isotropic, incompressible, hydrodynamic turbulence follows the Kolmogorov scaling relation, based on which non-thermal velocity dispersion ($\sigma_{nth}$) varies with the length-scale ($L$) with a power-law index ($p$) of 1/3 (Frisch 1995; Kolmogorov 1941). However, ISM is compressible as well as inhomogeneous and anisotropic; thus, $p$ may deviate from 1/3. There are also several theoretical studies exist in the literature based on the magnetized ISM, where it has been asserted that depending on the strength of the turbulence (strong and weak turbulences in comparison with the magnetic field), $p$ differs from the standard Kolmogorov scaling (Goldreich & Sridhar 1995; Sridhar & Goldreich 1994).

Turbulence measurements have been carried out in neutral ISM over the last several decades. For example, Larson (1979) studied the turbulence in neutral ISM and found that the power-law index ($p$) is 0.37. Likewise, using the electron density power spectrum, Armstrong et al. (1995) showed that the power-law index of the density power spectrum in the warm neutral medium is -11/3. Thereafter, several observational studies have indicated that there are variations of turbulence within neutral ISM (Choudhuri & Roy 2019; Kalberla & Haud 2019; Blagrave et al. 2017; Kalberla et al. 2017; Hennebelle & Falgarone 2012; Miville-Deschênes et al. 2003). For example, Choudhuri & Roy (2019), using the HI-21 emission spectra, demonstrated that the power-law is steeper in the velocity channels (observed towards a single line-of-sight), which are dominated by CNM clouds. Likewise, Kalberla & Haud (2019) also indicated that the power-law is steepest in channels dominated by CNM clouds and containing a minimal amount of WNM clouds. In addition to these observational studies, several theoretical studies (Xu *et al.* 2019; Kowal & Lazarian 2007; and references therein) have also shown that power-law index varies depending on the Mach number in the system. Accordingly, all of these studies indicate that a single power-law is unlikely to exist in the neutral ISM.

**Author for correspondence:** atanuphysics15@gmail.com
**Cite this article:**

Therefore, in this study, we examine whether a single power-law holds throughout the entire dataset, and if so, how it differs from the standard Kolmogorov scaling. Furthermore, we check whether this power-law is identical or not for different phases of neutral ISM.

In order to do this, we choose the prime tracer H I. The spectral width of H I-21 cm line does not arise from natural broadening; rather, it is caused due to thermal and non-thermal Doppler broadenings, both of which are Gaussian in shape (Carroll & Ostile 1996). The traditional way to obtain various physical properties through H I-21 cm spectral line study is to compare the emission and absorption spectra from the more or less same line-of-sight. Absorption spectra are obtained in the direction of background quasars. In contrast, emission spectra are observed from the nearby lines-of-sight by assuming that all the physical parameters are the same in emission and absorption (Murray *et al.* 2021; Roy *et al.* 2013*b,a*). After decomposing the emission and absorption spectra into multi-Gaussian components, one can obtain the following essential parameters for each of the components: column density ($N$), peak optical depth ($\tau_{peak}$), peak brightness temperature ($T_{B,peak}$), spin temperature ($T_s$), center velocity ($v_c$), and total velocity dispersion ($\sigma_{total}$ or upper limit of kinetic temperature $T_{k,max}$ )(Murray *et al.* 2018; Patra *et al.* 2018; Stanimirović *et al.* 2014; Mohan *et al.* 2004; Heiles & Troland 2003*b,a*). $T_s$ basically determines the relative level population of the two hyperfine levels of H I-21 cm line. In CNM, due to the sufficient collision, $T_s$ is strongly coupled with the system, and is equal to $T_k$. However, in other phases (UNM and WNM) where density is relatively low, $T_s$ is generally less than $T_k$, except in special environment, where the effect (Wouthuysen–Field effect) of strong Lyman-$\alpha$ photon is immense (Seon & Kim 2020; Field 1958).

In the study of the correlation between the $\sigma_{nth}$, and the $L$ (length-scale along the line-of-sight), it is necessary to obtain the information regarding the $T_k$ and the $P_{th}$ of each component. Since the line width of a component is the sum of a thermal and turbulence broadenings; therefore, to obtain the $\sigma_{nth}$, we need to know the $T_k$ and subtract the thermal velocity dispersion ($\sigma_{th}$) from the $\sigma_{total}$ of the line. In the same way, $P_{th}$ is needed for the derivation of $L$. Even though existing observational study of the thermal pressure in neutral ISM provides information on the thermal pressure (Jenkins & Tripp 2011), it is not possible from the observed emission-absorption study to converse from $T_s$ to $T_k$ unless some numerical model is assumed. This is particularly true for the UNM and the WNM phases, where $T_s$ is generally less than $T_k$ (Liszt 2001). Therefore, one has to first consider some numerical model for converting the $T_s$ of each of the components into the corresponding $T_k$ and then has to use this $T_k$ for calculating $\sigma_{nth}$ and $L$. Such numerical study has been performed by Liszt (2001) for neutral ISM. These aforementioned works provide us with the value of $P_{th}$ and the relationship between $T_s$ and $T_k$ which we use in our analysis to study the correlation between $\sigma_{nth}$ and $L$.

Section 2 describes the acquisition of the published H I emission-absorption data, including various measured and fitted parameters. In Section 3, we examine the relationship between $\sigma_{nth}$ and the corresponding $L$. In Section 4, we discuss different plausible reasons for breaking the power-law on a small scale as well study the properties of these components where power-law is not significantly observed. At last, in Section 5, we summarize our main conclusions.

## 2. H I emission-absorption data

All the fitted and measured parameters of H I emission-absorption spectra are taken from the published large-scale surveys as well as previous H I emission-absorption studies. We have used the data of large scale surveys: *Millennium* (hereafter HT03) (Heiles & Troland 2003*b,a*) and *21-SPONGE* (hereafter CEM18) (Murray *et al.* 2018) H I surveys. Likewise, we have also taken data from the work of Stanimirović *et al.* (2014) (hereafter SS14) or from the study where a single line-of-sight is observed (Patra *et al.* 2018). Note that we count only once the lines-of-sight, which are common in these surveys. For example, 22 out of 78 lines-of-sight in the HT03 have also been observed in the CEM18 survey. Therefore, we only take 56 lines-of-sight out of a total of 78 lines-of-sight from the HT03 survey. Additional restrictions have also been imposed in terms of the value of the $T_s$. We have not taken into consideration those components where $T_s$ is greater than the $T_{k,max}$ or the value of $T_s$ is only the lower limit of the actual value. Note that the former is not physically reliable, while the latter is not useful for our analysis. In the HT03 survey, both emission and absorption spectra were observed with Arecibo single-dish telescope. In contrast, in the CEM18 survey, absorption spectra were observed from the Very Large Array (VLA) telescope, and emission spectra were observed with the Arecibo single-dish telescope and Effelsberg 100-meter radio telescope (Effelsberg-Bonn H I survey by Winkel *et al.* (2016)). We would like to point out that since the absorption spectra obtained from interferometric observations are superior to those obtained from single-dish observations, we have used the components from the CEM18 survey which also observe in the HT03 survey. Another survey of H I emission-absorption study performed by Stanimirović *et al.* (2014) was made with the Arecibo single-dish telescope in the vicinity of Perseus molecular cloud. From this study, we take 19 sources out of 26 sources. 3 sources have been rejected because these sources were also observed in other surveys. Another 4 sources are toward the Perseus molecular cloud. Thus, there is a high possibility of detecting the gas components associated with the Perseus molecular cloud in addition to the general diffuse ISM gas components (see the Fig. 1 of Stanimirović *et al.* (2014)). We have also used gas component studied by Patra *et al.* (2018), where H I emission-absorption spectra were observed for a particular line-of-sight; absorption spectra were observed with Giant Metrewave Radio Telescope (GMRT), and emission spectra were obtained from the Parkes Radio telescope for comparison. This work, we have taken into consideration, albeit observed towards a single line-of-sight, because here the successfully recovered component's $T_s$ is ∼ 1000 K. This kind of broad weak component is rarely detected in absorption due to the low optical depth ($\tau_v$) of the line. Velocity resolution of the CEM18 and HT03 surveys was ∼ 0.40 km sec⁻¹, whereas it was ∼ 0.16 km sec⁻¹ for the SS14 survey. Decomposition of the emission spectra, in general, is complicated due to the unknown relative positions of the gas clouds, self-absorption as well as absorption effects of the front clouds on the back clouds. By using different permutations of the gas clouds' positions, the assumption of the unknown filling factor (*f*), and with the help of decomposed absorption spectra, emission profiles have been decomposed into multi-Gaussian components in these surveys. For a detailed discussion about the decomposition of emission and absorption spectra, see the papers of Murray *et al.* (2018); Stanimirović *et al.* (2014); Roy *et al.* (2013b,a); Heiles & Troland (2003b,a). Number of gas components toward different background sources, that we have used for our analysis are mentioned in the Appendix A. After adding all these gas clouds, we have total 378 components for studying the correlation between $\sigma_{nth}$ and $L$. Among the 378 components, 196 components are taken from the CEM18 survey, 113 components come from the HT03 survey, 68 gas clouds come from the SS14 survey, and the remaining component is taken from the work of Patra *et al.* (2018).

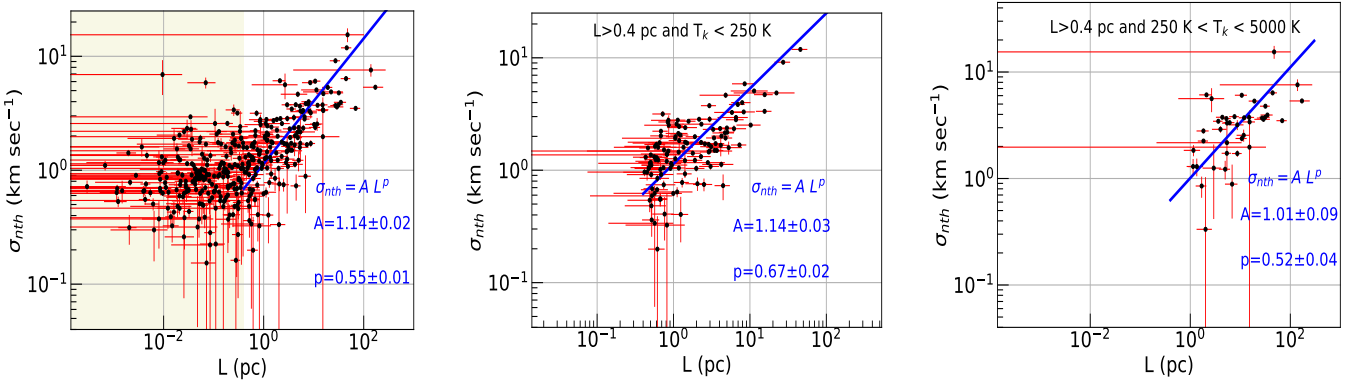

**Figure 1:** Left: Correlation between the non-thermal velocity dispersion ($\sigma_{nth}$) and the length-scale ($L$) for the whole data set. For the case where $L$ is > 0.40 pc (see Appendix C for details), we fit this dataset with a power-law ($AL^p$) using Bayesian statistics (see Appendix D). The most probable values for $A$ and $p$ are 1.14 and 0.55 respectively. The fitted line, which is shown in blue solid color is made with these values. Here the yellow shaded area covers the entire dataset with $L < 0.40$ pc. Middle: Same correlation between the non-thermal velocity dispersion ($\sigma_{nth}$) and the length-scale ($L$) for the data set where $L$ is > 0.40 pc and $T_k$ is < 250 K. Here also the fitted line is made with the most probable values $A$ and $p$, obtained from the Bayesian statistics. These values of $A$ and $p$ are 1.14 and 0.67 respectively. Right: Same correlation as left and middle but for the dataset where $L$ is > 0.40 pc and $T_k$ is between 250 K and 5000 K. Here the most probable values of $A$ and $p$ are 1.01 and 0.52 respectively. Here also the fitted line is drawn with these values.

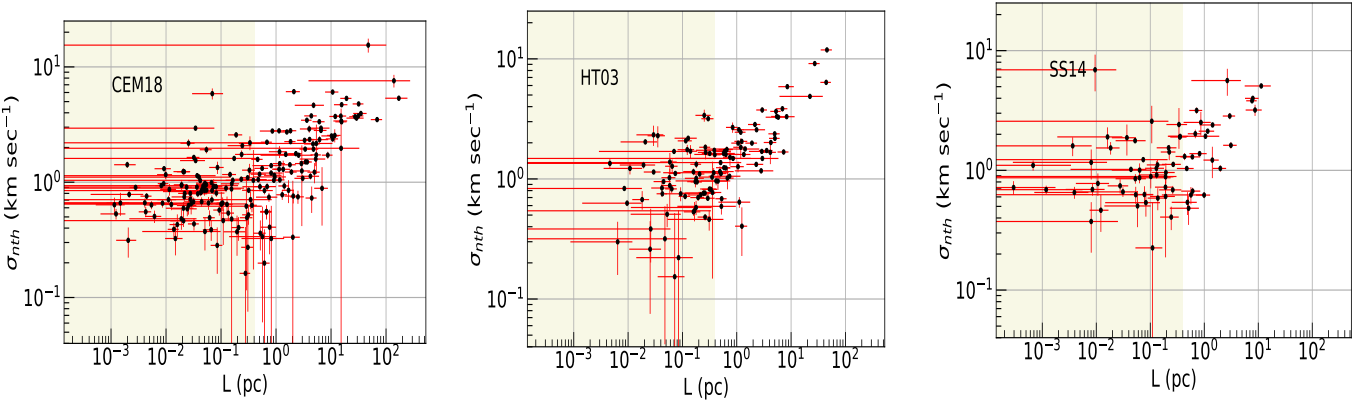

**Figure 2:** Left: Correlation between the $\sigma_{nth}$ and the $L$ for the CEM18 survey. Middle: Same correlation but for the HT03 survey (only for those components which have taken from this survey). Right: Same correlation but for the SS14 survey (only for those components which have taken from this survey). In all three figures, the yellow shaded area corresponds to the same area as shown in Figure 1.

## 3. Correlation between the non-thermal velocity dispersion and the length-scale

To analyze the correlation between $\sigma_{nth}$ and $L$, it is necessary to calculate these parameters for each gas component. $\sigma_{nth}$ is obtained by the formula:

$$\sigma_{nth} = \sqrt{\sigma_{total}^2 - \sigma_{th}^2} \qquad (1)$$

Here, $\sigma_{total}$ is equal to $(T_{k,max}/121)^{\frac{1}{2}}$ and $\sigma_{th}$ is equal to $(T_k/121)^{\frac{1}{2}}$. $\sigma_{total}$ is directly obtained from the observed emission-absorption measurement. On the other-hand, for the $\sigma_{th}$, we consider the numerical model of Liszt (2001). This has led us to convert each component's $T_s$ into its $T_k$ and measure the $\sigma_{nth}$. Likewise, $L$ can be calculated for each component by the formula:

$$L = N(HI) T_k k_B / P_{th} \qquad (2)$$

$N(HI)$ of each component is obtained from the joint emission-absorption study. The value of $P_{th}$ is obtained from the observational work of Jenkins & Tripp (2011). According to their analysis, the observed thermal pressure profile of the neutral ISM exhibits a log-normal distribution with a median value of $\sim 3800$ K cm$^{-3}$. We take this value of $P_{th}$ for our analysis. Additionally, we note that from the numerical model of Liszt (2001), we convert the $T_s$ into $T_k$ for each component associated with that particular $P_{th}$.

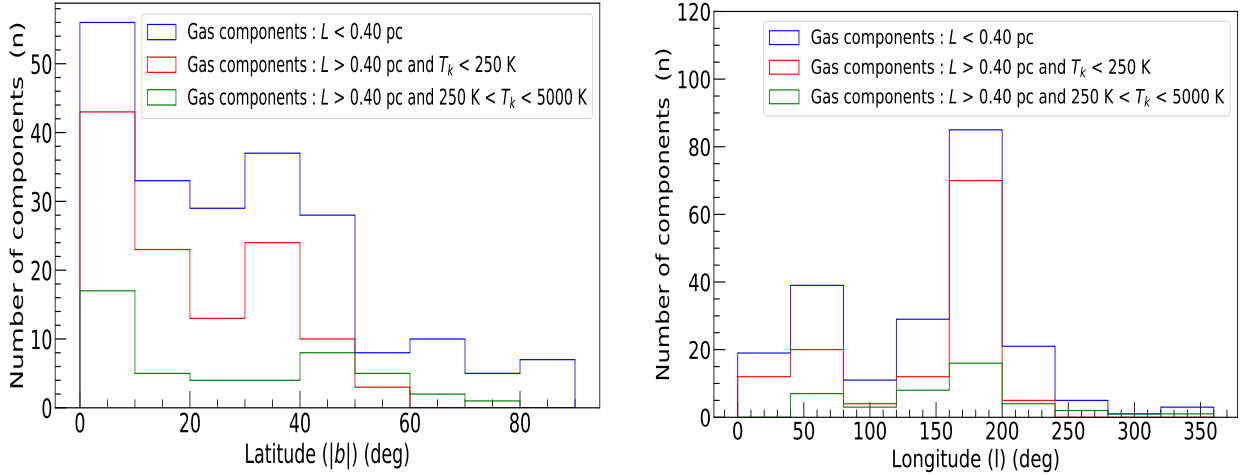

**Figure 3:** Left: Histogram plots of latitude ($|b|$) of three types of gas components. Components for which $L$ is < 0.40 pc is shown in blue color. Components for which $L$ is > 0.40 pc and $T_k$ is < 250 K is shown in red color. Lastly, components for which $L$ is > 0.40 pc and $T_k$ is between 250 K and 5000 K is shown in green color. Right: Same histogram plots but for the longitude ($l$).

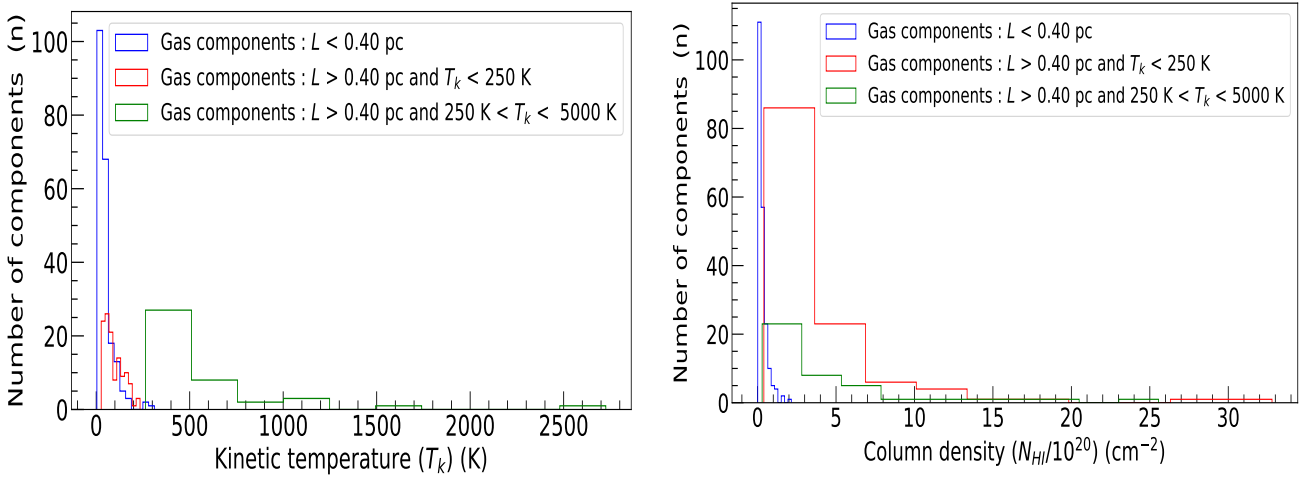

**Figure 4:** Left: Histogram plots of the kinetic temperature ($T_k$) of three types of gas components. Components for which $L$ is < 0.40 pc is shown in blue color. Components for which $L$ is > 0.40 pc and $T_k$ is < 250 K is shown in red color. Lastly, components for which $L$ is > 0.40 pc and $T_k$ is between 250 K and 5000 K is shown in green color. Right: Same histogram plots but for the column density ($N(HI)$).

Left panel of Fig. 1 shows the correlation between $\sigma_{nth}$ and $L$ for the entire dataset. In this figure, we also include the $1\sigma$ errors of $L$ and $\sigma_{nth}$. Detailed information regarding how we calculate the errors of $L$ and $\sigma_{nth}$ is provided in Appendix B. Visually, from the plot it appears that there is no significant correlation between $\sigma_{nth}$ and $L$ at the lower length scales. However, to verify this, we have calculated the Spearman correlation coefficient ($S$) for different bins of the dataset and identified the particular length scale ($L$) at 0.40 pc, below which there is no significant correlation exists. Detail about this is discussed in Appendix C. Now, we fit a power-law $AL^p$ to dataset that exhibits $L$ > 0.40 pc. Please note that, we use the standard Markov chain Monte Carlo (MCMC) method and obtain the maximum probable values and the parameter spaces of $A$ and $p$. Detailed discussion regarding the fitting is discussed in Appendix D. From this method, we obtain the most probable values of $A$ and $p$ are 1.14 and 0.55 respectively and the standard deviations of these quantities are 0.02 and 0.01 respectively. We note that even if we consider all possible ranges (99.7% probability of getting the value within this range) of $p$, the value of $p$ is quite different from the standard Kolmogorov scaling, where the value of $p$ is 0.33. The main reason for this difference is that the dataset is dominated by CNM clouds; where the turbulence is supersonic and supported by the sonic Mach number ($M_s$) distributions of the clouds (see the Appendix E for this). A better understanding of this argument can be obtained by separating the good fitted dataset (where $L$ > 0.40 pc) into two groups: one for which $T_k$ is less than 250 K, and the other for which $T_k$ is greater than 250 K. As mentioned in the introduction section, the components with a $T_k$ below 250 K are definitely in the CNM phase. Likewise, components with a $T_k$ of more than 250 K but less than 5000 K are considered to be in the UNM phase. Lastly, components with a $T_k$ greater than 5000 K are believed to be in the WNM phase. We notice that in the dataset, the maximum $T_k$ for the gas components is 2725 K. Therefore, all the gas components above 200 K fall into the UNM phase. It is noteworthy that in our analysis, we found a total of 213 gas components with $L$ < 0.40 pc, 113 components with $L$ > 0.40 pc and $T_k$ < 200 K, and 42 gas components with $L$ > 0.40 pc and 250 < $T_k$ < 5000 K.

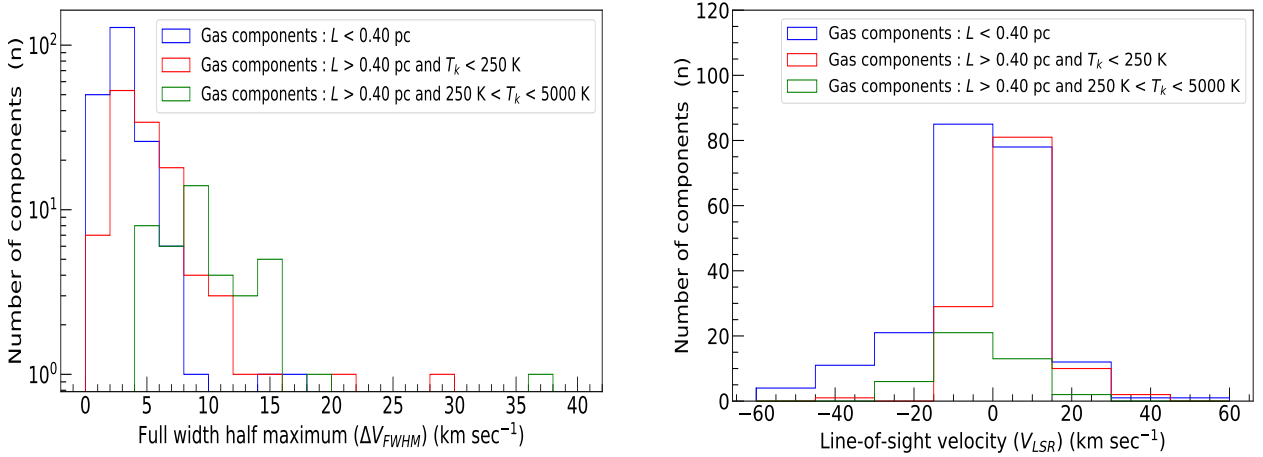

**Figure 5:** Left: Histogram plots of the full width at half maxima ($\Delta V_{FWHM}$) of three types of gas components. Components for which $L$ is < 0.40 pc is shown in blue color. Components for which $L$ is > 0.40 pc and $T_k$ is < 250 K is shown in red color. Lastly, components for which $L$ is > 0.40 pc and $T_k$ is between 250 K and 5000 K is shown in green color. Right: Same histogram plots but for the line-of-sight velocity ($V_{LSR}$).

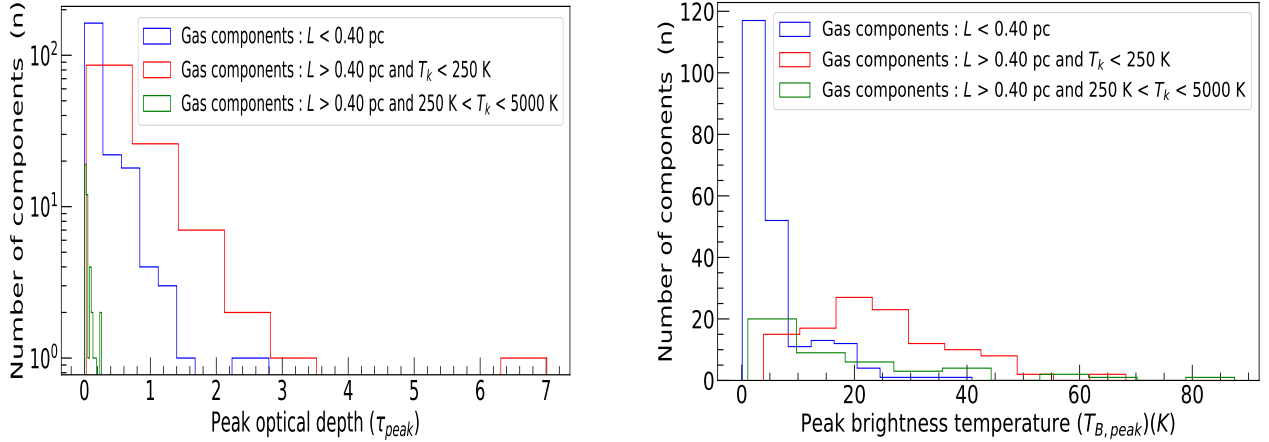

**Figure 6:** Left: Histogram plots of the peak optical depth ($\tau_{peak}$) of three types of gas components. Components for which $L$ is < 0.40 pc, which is shown in blue color. Components for which $L$ is > 0.40 pc and $T_k$ is < 250 K is shown in red color. Lastly, components for which $L$ is > 0.40 pc and $T_k$ is between 250 K and 5000 K is shown in green color. Right: Same histogram plots but for the peak brightness temperature ($T_{B,peak}$).

In the middle and right panels of Fig. 1, we have plotted the same correlation for the CNM and UNM phases respectively. Note that we have rejected all the components for which the measured $L$ is < 0.40 pc. The same MCMC method is used to obtain the most probable values as well as the parameter spaces of $A$ and $p$.

All the details are mentioned in Appendix D. The most probable values of $A$ and $p$ that we obtain are 1.14 and 0.67 respectively for the CNM clouds and for the UNM clouds these are 1.01 and 0.52 respectively. Additionally, 99.7% of the probabilities of getting the values of $A$ and $p$ in the CNM phase are between 1.05 and 1.23, and 0.58 and 0.76, respectively. However, in the UNM phase, these values vary from 0.74 to 1.28 and from 0.40 to 0.64, respectively. Here we would like to mention that, in the UNM phase, most of the clouds reside below 500 K (see the left panel of Fig. 4). Therefore, for well constrained results of $A$ and $p$, it is necessary to obtain a substantial number of data points that cover the entire range of the UNM phase. However, based on the possible values of $p$, it is evident that in the CNM phase power-law is steeper and significantly differs from the Kolmogorov scaling. It is because, in the CNM phase, shocks are frequently formed (Audit & Hennebelle 2005). Therefore, a significant amount of energy is dissipated on a relatively large scale, which causes the power-law to become steeper (see also the Appendix E for the sonic Mach number ($M_s$) distribution of the gas components).

## 4. Plausible reasons for breaking the power-law

Despite $\sigma_{nth}$ exhibiting a strong correlation with $L$ above 0.40 pc, it does not show any significant correlation with $L$ below 0.40 pc. For each individual survey, we have plotted $\sigma_{nth}$ vs. $L$ in the left, middle, and right panels of Fig. 2 . These have been plotted separately to determine whether the same trend is observed across all of these surveys. We notice that there is a similar trend in all of these surveys, although the resolutions of the observations differ (see the Section 2). Therefore, it is interesting to know about the properties of these small length-scale components ($L < 0.40$ pc) e.g., $T_k$, $N(HI)$, $\tau_{peak}$, $T_B$, $V_{LSR}$, etc. Furthermore, it is important to compare the properties of these gas clouds with those of other gas components, which exhibit a power-law relationship between $\sigma_{nth}$ and $L$.

In the left and right panels of Fig. 3, we present the histogram plots of the latitude ($|b|$) and longitude ($l$) of all these three types of components. The components for which $L$ is < 0.40 pc is shown by the blue line. Components where $L$ is > 0.40 pc and $T_k$ is < 250 K is indicated by the red line. In contrast, components where $L$ is > 0.40 pc and $T_k$ is between 250 K and 5000 K is indicated by the green line. We use the same colors for the following comparison plots as well. From these figures, we see that all three types of components are widely distributed throughout the Galaxy. There is no spatial latitude or longitude found where these low length-scale ($L < 0.40$ pc) components reside.

In the left and right panels of Fig. 4, we show the histogram plots of $T_k$ and $N(HI)$ of these three types of components. From the left panel, it indicates that the median value of $T_k$ of these low length-scale ($L < 0.40$ pc) components is lower in comparison to the other gas clouds. The median value of $T_k$ for the former is ∼ 30 K, whereas for the latter are ∼ 75 K and ∼ 434 K respectively. In the right panel of the same figure, we compare the column density distributions of these gas components. The median value of the low length-scale gas components ($L < 0.40$ pc) is ∼ 2.0 × $10^{19}$ cm$^{-2}$, which is one order lower compared to the other gas clouds, where the median value of the $N(HI)$ is ∼ 2.2 × $10^{20}$ cm$^{-2}$.

Left and right panels of Fig. 5 shows the histogram plots of $\Delta V_{FWHM}$ and $V_{LSR}$ of the gas components. From the left panel, median value of $\Delta V_{FWHM}$ of the low length-scale components is ∼ 2.4 km sec$^{-1}$ which corresponds to $T_{k,max}$ = 126 K, whereas these are ∼ 4.1 km sec$^{-1}$ and ∼ 9.3 km sec$^{-1}$ for the other gas clouds. Likewise, from the right panel, median value of $V_{LSR}$ of the low length-scale components is ∼ - 1.5 km sec$^{-1}$, whereas these are ∼ + 5.0 km sec$^{-1}$ and ∼ - 3.4 km sec$^{-1}$ for the other gas clouds respectively. we notice that the line-of-sight velocity $V_{LSR}$ of the low length-scale components ($L < 0.40$ pc) has a nearly symmetric distribution around 0 km sec$^{-1}$. If these components are at a very high Galactic height (> 1 kpc, then the work of Begum et al. (2010b,a)), then Galactic rotation will be negligible and the line-of-sight velocity ($V_{LSR}$) will be skewed toward negative velocity. As this is contrary to our result, we can consider that these clouds do not reside at a very high Galactic height.

In the left and right panels of Fig. 6, we show the histogram plots of $\tau_{peak}$ and $T_{B,peak}$ of the gas components. From the left panel, median value of $\tau_{peak}$ of the low length-scale components is ∼ 0.09, whereas these are ∼ 0.37 and ∼ 0.03 for the other gas clouds. Likewise, from the right panel, median value of $T_{B,peak}$ of the low length-scale components is ∼ 4 K, whereas these are ∼ 23 K and ∼ 13 K for the other gas clouds. From all these plots and the comparison with other gas clouds, it is evident that these small length-scale components ($L < 0.40$ pc ) are mostly cold and can be thought of as a special class in the CNM phase.

Now the reason for breaking the power-law for these small length-scale ($L < 0.40$ pc) clouds can be described in two ways. First, these small cold components may be formed due to the stellar outflows, winds, etc (Braun & Kanekar 2005). It may also possible that the properties of the turbulence is different there and cannot be fit with a simple power-law. Another possibility is slightly different. It is related to the rough thermal pressure ($P_{th}$) equilibrium of all the gas components. From the thermal stability analysis, it has been shown that various phases of neutral ISM are roughly in thermal pressure equilibrium ($P_{th}$). It may possible that these components reside at a lower thermal pressure ($P_{th}$). Stanimirović & Heiles (2005) discussed in great detail the different possible scenarios regarding these small gas cloud components.

Here we also note that these types of small cloud components have already been obtained in the Galactic Arecibo L-band Feed Array H I (GALFA-H I) survey (Begum et al. 2010b,a), where the beam size of the telescope was 3.5 arcmin and velocity resolution was 0.18 km sec$^{-1}$. They found a total of 96 components within the searched area of ∼ 4600 deg$^2$. It is interesting to note that the properties of these clouds are more or less similar to our results. Depending on the distance, they argued two possible scenarios. If the distances of the clouds are < 1 kpc, then these clouds have number density ($n$) ∼ 1 cm$^{-3}$ and $P_{th}$ is ≤ 100 K cm$^{-3}$. These clouds are similar to the H I gas components found in the disk-halo interface region believed to be originated from the expelled of hot gas from the Galactic disk by supper bubbles. On the other hand, if the clouds are within 100 pc, these clouds are mostly sub-parsec clouds with $n$ ∼ 10 cm$^{-3}$ and $P_{th}$ ∼ 3000 K cm$^{-3}$. These clouds are formed due to stellar outflow, radiation, etc.

## 5. Conclusions

This paper studies the correlation between the non-thermal velocity dispersion ($\sigma_{nth}$) and the length-scale ($L$) for a large number of H I gas components taken from various published surveys. Our main conclusions from this study are as follows:

(1) We observe that there is a power-law relationship between non-thermal velocity dispersion ($\sigma_{nth}$) and the length-scale ($L$) above $L$ of 0.40 pc. However, there is a break in the power-law below 0.40 pc, where $\sigma_{nth}$ does not show any significant correlation with $L$. The most probable values of $A$ and $p$ for the length-scale ($L$) > 0.40 pc, that we obtain from Markov chain Monte Carlo method (MCMC) is 1.14 and 0.55 respectively. we also observe that $A$ and $p$ values (99.7% probability of getting the value within this range) vary between 1.08 and 1.20 and between 0.52 and 0.58, respectively. From the possible value of $p$, we can argue that power-law is steeper than the standard Kolmogorov scaling. The main reason for this is that the dataset is dominated by the CNM clouds. This is even more clear if we divide the good fitted data points ($L > 0.40$ pc) into two categories: one for which $T_k$ is less than 250 K, and the other for which $T_k$ is greater than 250 K. The same MCMC method is used to obtain the most probable values as well as the parameter spaces of $A$ and $p$. We obtain the most probable values of $A$ and $p$ are 1.14 and 0.67 respectively for the CNM clouds and for the UNM clouds we obtain 1.01 and 0.52 respectively. The values of $A$ and $p$ ( 99.7% probability of getting the value within this range) can vary from 1.05 to 1.23 and from 0.58 to 0.76, respectively for the CNM gas clouds. Conversely, these values can vary between 0.74 to 1.28 and and between 0.40 to 0.64 respectively, in the UNM gas clouds. Most of the clouds in the UNM phase lie below 500 K. Therefore, to obtain well-constrained results for $A$ and $p$, more data points are required in order to cover the full range of the UNM phase. However, based on the most probable value and the possible ranges of $p$, we can conclude that in CNM phase, the power-law is steeper and significantly differs from the standard Kolmogorov scaling, which can be attributed to a shock-dominated medium.

(2) We discuss the properties of these small length-scale ($L < 0.40$ pc) clouds where there is no significant correlation has been observed and compare with the other gas components. We find that these small length-scale ($L < 0.40$ pc) clouds are widely spread across the Galaxy and the line-of-sight velocity ($V_{LSR}$) has a nearly symmetric distribution around 0 km sec$^{-1}$. Median values of $T_k$ and ($NHI$) these components are $\sim 30$ K and $\sim 2.0 \times 10^{19}$ cm$^{-2}$ respectively. Likewise, the median values of peak optical depth ($\tau_{peak}$) and the peak brightness temperature ($T_{B,peak}$) of these clouds are 0.09 and $\sim 4$ K respectively. These clouds are mostly cold and can be considered as a special class in the cold neutral medium (CNM).

**Acknowledgements.** We thank Harvey Liszt for providing his numerical results to us. AK gratefully acknowledges support from ANID BASAL project FB210003.

# References

Armstrong, J. W. & Rickett, B. J. & Spangler S. R. 1995, ApJ, 443, 209

Audit, E. & Hennebelle, P. 2005, A&A, 433, 1

Begum, A., Stanimirović, S., Peek, J. E., et al. 2010a, ApJ, 722, 395

Begum, A., Stanimirović, S., Peek, J. E. G., et al. 2010b, in Astronomical Society of the Pacific Conference Series, Vol. 438, The Dynamic Interstellar Medium: A Celebration of the Canadian Galactic Plane Survey, ed. R. Kothes, T. L. Landecker, & A. G. Willis, 126

Blagrave, K. et al. 2017, ApJ, 834, 126

Braun, R. & Kanekar, N. 2005, A&A, 436, L53

Carroll, B. W. & Ostlie, D. A. 1996, An Introduction to Modern Astrophysics

Choudhuri, S. & Roy, N. 2019, MNRAS, 483, 3437

Field, G. B. 1958, Proceedings of the IRE, 46, 240

Field, G. B. 1965, ApJ, 142, 531

Field, G. B., Goldsmith, D. W., & Habing, H. J. 1969, ApJ, 155, L149

Frisch, U. 1995, Turbulence. The legacy of A.N. Kolmogorov

Goldreich, P. & Sridhar, S. 1995, ApJ, 438, 763

Heiles, C. & Troland, T. H. 2003a, ApJS, 145, 329

Heiles, C. & Troland, T. H. 2003b, ApJ, 586, 1067

Hennebelle, A. & Falgarone, E. 2012, A& ARv, 20, 55

Jenkins, E. B. & Tripp, T. M. 2011, ApJ, 734, 65

Kalberla, P. M. W. & Haud, U. 2019, A& a, 627, A112

Kalberla, P. M. W. et al., 2017, A& A, 607, A15

Kolmogorov, A. N. 1941, Akademiia Nauk SSSR Doklady, 32, 16

Kowal, G. & Lazarian, A. 2007, ApJL, 666, L69

Larson, R. B. 1979, MNRAS, 186, 479

Liszt, H. 2001, A&A, 371, 698

Miville-Deschênes M.-A. et al. 2001, A&A, 411, 109

Mohan, R., Dwarakanath, K. S., & Srinivasan, G. 2004, Journal of Astrophysics and Astronomy, 25, 143

Murray, C. E., Stanimirović, S., Goss, W. M., et al. 2015, ApJ, 804, 89

Murray, C. E., Stanimirović, S., Goss, W. M., et al. 2018, ApJS, 238, 14

Murray, C. E., Stanimirović, S., Heiles, C., et al. 2021, ApJS, 256, 37

Patra, N. N., Kanekar, N., Chengalur, J. N., & Roy, N. 2018, MNRAS, 479, L7

Roelfs, M. & Kroon, P. C. 2022, tBuLi/symfit: symfit 0.5.5, Zenodo

Roy, N., Kanekar, N., Braun, R., & Chengalur, J. N. 2013a, MNRAS, 436, 2352

Roy, N., Kanekar, N., & Chengalur, J. N. 2013b, MNRAS, 436, 2366

Sridhar, S. & Goldreich, P. 1994, ApJ, 432, 612

Stanimirović, S. & Helies, C. 2005, ApJ, 631, 371

Stanimirović, S., Murray, C. E., Lee, M.-Y., Heiles, C., & Miller, J. 2014, ApJ, 793, 132

Seon, K.-i. & Kim, C.-G. 2020, ApJS, 250, 9

Winkel, B., Kerp, J., Flöer, L., et al. 2016, A&A, 585, A41

Wolfire, M. G., Hollenbach, D., McKee, C. F., Tielens, A. G. G. M., & Bakes, E. L. O. 1995, ApJ, 443, 152

Wolfire, M. G., McKee, C. F., Hollenbach, D., & Tielens, A. G. G. M. 2003, ApJ, 587, 278

Xu, S., Ji, S., & Lazarian, A. 2019, ApJ, 878, 157

## Appendix A. Acquisition of the H I emission-absorption data

In Table 2, we describe the gas components that we have used for our analysis. The first column lists all the background sources. Columns 2 and 3 represent the longitude and latitude of the sources. Columns 4, 5, 6, and 7 indicate the peak optical depth ($\tau_{peak}$), full width at maximum (FWHM), spin temperature ($T_s$), and kinetic temperature ($T_k$) of these gas components as well as their corresponding errors. Column densities of the errors are mentioned in column 8. Columns 9 and 10 describe the $\sigma_{nth}$ and the length-scale ($L$) of the gas clouds and their respective errors. Lastly, column 11 denotes the references from where we have taken the data.

## Appendix B. Error calculations of $\sigma_{nth}$ and $L$

In this analysis, errors arise from both the length-scale ($L$) and the non-thermal broadening ($\sigma_{nth}$). $L$ is obtained by the formula:

$$L = \frac{N(HI)T_k}{P_{th}} = \frac{c.T_s.\tau_{peak}.V_{FWHM}.T_k}{P_{th}} \qquad (3)$$

Here, $N(HI)$ is the column density of the H I gas, $T_s$ is the spin temperature of the gas, $\tau_{peak}$ is the peak optical depth, $V_{FWHM}$ is full width at half maximum of the spectrum, $T_k$ is the kinetic temperature, $P_{th}$ is the thermal pressure, and $c$ is the constant. The majority of the errors are uncorrelated except in the cases of $T_s$ and $T_k$. They are correlated with one another. Therefore, the error of $L$ is obtained by the following equation:

$$\Delta L \approx L \sqrt{\left(\frac{\Delta T_s}{T_s} + \frac{\Delta T_k}{T_k}\right)^2 + \left(\frac{\Delta \tau_{peak}}{\tau_{peak}}\right)^2 + \left(\frac{\Delta V_{FWHM}}{V_{FWHM}}\right)^2 + \left(\frac{\Delta P_{th}}{P_{th}}\right)^2} \qquad (4)$$

We note that errors of different parameters have been mentioned in Table 2, except for the thermal pressure. In this case, we take the fixed value of $\Delta P_{th} = 800$ K cm$^{-3}$ after inspecting the log-normal profile of thermal pressure profile in neutral ISM from the work of Jenkins & Tripp (2011). We would also like to mention that measurement of $T_s$ and its error taken from these published surveys are obtained not through least-square fitting but rather by permuting the positions of all gas clouds, taking into account absorption effects, and weighting each combination according to their similarities to the observed emission spectra. Therefore, the error of $T_s$ is generally higher than the errors of other quantities obtained by least-square fitting.

In the same way, we also obtain the $\sigma_{nth}$ by the formula:

$$\sigma_{nth} = \sqrt{\sigma_{total}^2 - \sigma_{th}^2} \qquad (5)$$

We denote the error of $\sigma_{total}$ by $\Delta\sigma_{total}$ and the error of $\sigma_{th}$ by $\Delta\sigma_{th}$. Now, the error of $\sigma_{total}^2$, if we refer to as $\Delta p$, is related to $\Delta\sigma_{total}$ by:

$$\Delta p = 2.\sigma_{total}.\Delta\sigma_{total} \qquad (6)$$

On the other hand, $\sigma_{th}^2$ is related to $T_k$ by the following equation:

$$\sigma_{th}^2 = \frac{T_k}{a} \qquad (7)$$

If we denote the error of $\sigma_{th}^2$ as $\Delta q$, then it is related to $\Delta T_k$ by the formula:

$$\Delta q = \frac{\Delta T_k}{a} \qquad (8)$$

Now, the error of $\sigma_{nth}^2$, if we describe it by $\Delta r$, then it is related to $\Delta p$ and $\Delta q$ by:

$$\Delta r = \sqrt{(\Delta p)^2 + (\Delta q)^2} \qquad (9)$$

or,

$$\Delta r = \sqrt{\left(4.\sigma_{total}^2.\Delta\sigma_{total}^2 + \frac{\Delta T_k^2}{a^2}\right)} \qquad (10)$$

Finally, the error of $\sigma_{nth}$ is obtaind by the formula:

$$\sigma_{nth} = \frac{\Delta r}{2.\sigma_{nth}} = \frac{\sqrt{\left(4.\sigma_{total}^2.\Delta\sigma_{total}^2 + \frac{\Delta T_k^2}{a^2}\right)}}{2.\sqrt{\left(\sigma_{total}^2 - \frac{T_k}{a}\right)}} \qquad (11)$$

## Appendix C. Calculation of the length-scale cutoff using Monte Carlo simulation

We determine the particular length-scale up to which the maximum correlation between $\sigma_{nth}$ and $L$ is obtained by calculating the Spearman correlation coefficient ($S$) and the $p$-value. In our case, both $L$ and $\sigma_{nth}$ have errors. Therefore, there is no way to determine a particular value of Spearman correlation coefficient ($S$). That's why considering the error, we create $10^5$ random instances from the dataset and run the Monte Carlo simulation. We then finally obtain a distribution of Spearman correlation coefficient ($S$) and $p$-value. According to the length-scales, we have performed this procedure separately for different dataset and derived median values and $1\sigma$ errors of $S$ and $p$ values. Those results are mentioned in the Table 1. From these results we notice that the median value of $S$ is highest (0.64) in the case where $L$ is > 0.40 pc and lowest (0.13) for the case $L$ is < 0.40 pc. As an example, in the Figures 7 & 8, we have also shown the distribution of $S$ and $p$-values for the cases where $L$ is > 0.40 and $L$ is < 0.40. From the right panel of Figure 8, we can see that the value of $p$ is distributed from 0 to 1. This clearly indicates that there is no significant correlation exists where the value of $L$ is < 0.40 pc. We therefore fit the data with a power-law where $L$ is > 0.40 pc.

## Appendix D. Fitting the power-law using Bayesian statistics (Markov chain Monte Carlo method)

We fit the power-law using the Markov chain Monte Carlo (MCMC) method using the python package *Bilby*. For the purpose of fitting the power-law, we first convert $L$ and $\sigma_{nth}$ into logarithmic space. In addition, we also convert the errors of $L$ and $\sigma_{nth}$ in the logarithmic space. We then fit a linear function with the dataset. For converting the error in the logarithmic space, we use the formula: $\Delta(log x) = log\left(\frac{x + \frac{\Delta x}{2}}{x - \frac{\Delta x}{2}}\right)$. We provide the fitting function and the prior information about the model parameters. As inputs for the prior information of the model parameters, we provide the ranges of $A$ and $p$ from 0.6 to 2.0 and 0.1 to 0.9 uniformly. By taking these values and optimizing the likelihood, we obtain the posterior probabilities of the model parameters $A$ and $p$. We first do the analysis for the dataset where $L$ is > 0.40 pc. Then we perform this for the dataset where $L$ is > 0.40 pc and $T_k$ is < 250 K. And at last, we perform this for the gas clouds where $L$ is > 0.40 pc and $T_k$ is between 250 K and 5000 K. For the first case, we obtain the most probable values of $A$ and $p$ are 1.14 and 0.55 respectively. For the second case, the most probable values are 1.14 and 0.67 respectively and for the third case these are 1.01 and 0.52 respectively. Likewise, for the first case, the allowed values (99.7% probability of getting the values within this range) of $A$ and $p$ vary from 1.08 to 1.20 and 0.52 to 0.58 respectively. For the second case, these values are from 1.05 to 1.23 and from 0.58 to 0.76 respectively. And for the last case, these are from 0.74 to 1.28 and from 0.40 to 0.64 respectively. All these plots are protrayed in Figures 9 and 10.

## Appendix E. Sonic Mach number distribution of the gas components

In Figure 11, we have shown the sonic Mach number ($M_s$) distributions of the gas components for which $L$ is > 0.4 pc, and $L$ is > 0.4 pc and $T_k$ is < 250 K respectively. From both these plots, we notice that for most of the gas components, $M_s$ is > 1 and the median value of $M_s$ is ~ 4 in both cases, which suggests that, turbulence is supersonic. We obtain the value of $M_s$ using the following formula: $M_s = (\frac{\sigma_{nth,3D}}{c_s})$. The three dimensional non-thermal velocity dispersion $\sigma_{nth,3D}$, is calculated from the measured one dimensional non-thermal velocity dispersion ( $\sigma_{nth}$) after multiplying with $\sqrt{3}$, which comes under the assumption of isotropic nature of turbulence. Similarly, sound speed $c_s$ is obtained by the formula: $\sqrt{\frac{k_B T_k}{\mu m_H}}$. $k_B$ is the Boltzmann constant, $T_k$ is the kinetic temperature of the gas cloud, $m_H$ is the mass of the hydrogen atom, and $\mu$ is the mean molecular weight of the gas in neutral ISM, which we use 1.4 for our analysis (Murray *et al.* 2015; Heiles & Troland 2003b).

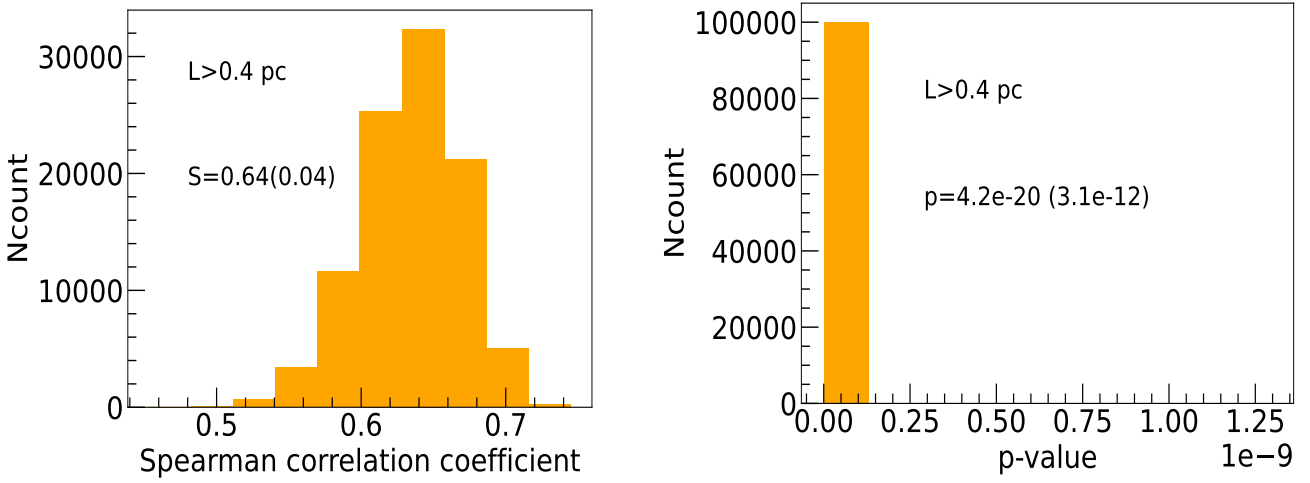

**Figure 7**: Left: Histogram plot of Spearman correlation coefficient ($S$) for the dataset where $L$ is > 0.40 pc. Median value of $S$ is 0.64 and the $1\sigma$ uncertaintity is 0.04. Right: Histogram plot of $p$-value for the dataset where $L$ is > 0.40 pc. Median value of $p$-value is 4.2e-20 and the $1\sigma$ uncertaintity is 3.1e-12.

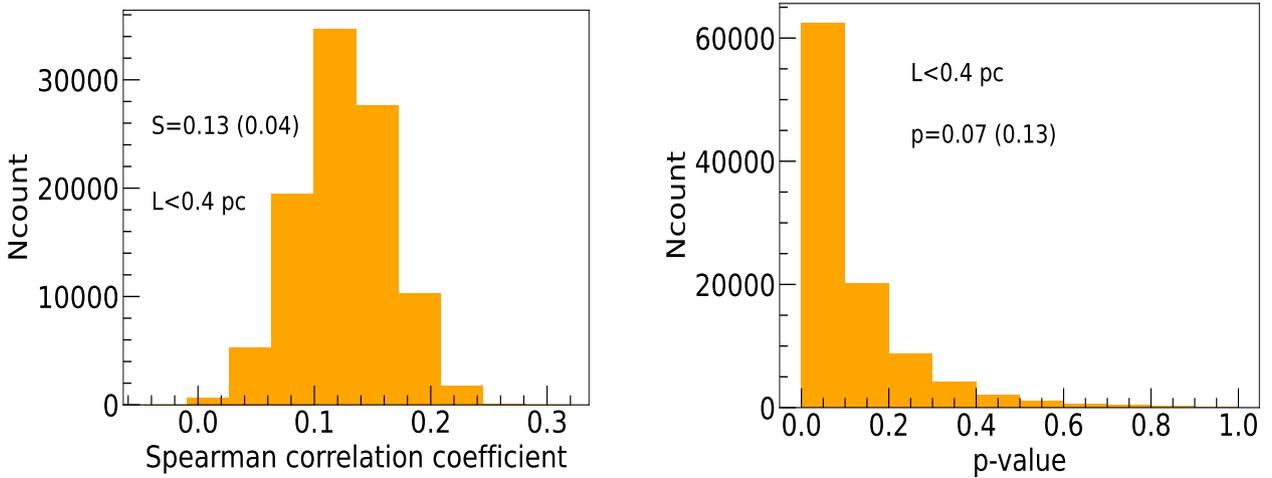

**Figure 8**: Left: Histogram plot of Spearman correlation coefficient ($S$) for the dataset where $L$ is < 0.40 pc. Median value of $S$ is 0.13 and the $1\sigma$ uncertaintity is 0.04. Right: Histogram plot of $p$-value for the dataset where $L$ is < 0.40 pc. Median value of $p$-value is 0.07 and the $1\sigma$ uncertaintity is 0.13.

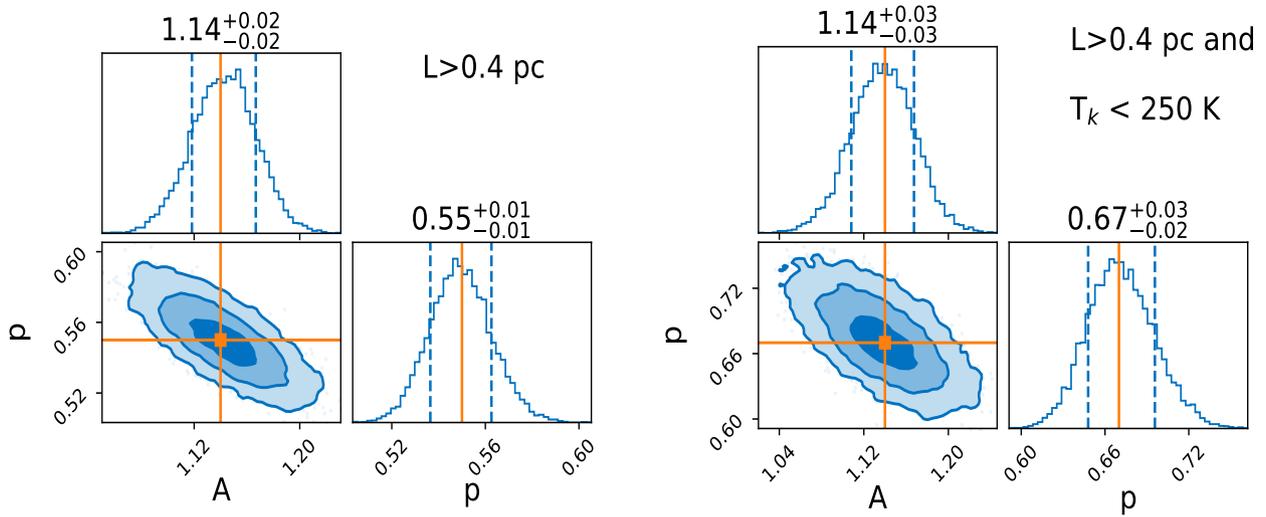

**Figure 9:** Left: Histogram plot and the phase space diagram of $A$ and $p$ for the data set where $L$ is > 0.40 pc. These are obtained from the Bayesian statistics. Most probable value of $A$ and $p$ are 1.14 and 0.55 respectively. Right: Same histogram plot and the phase space diagram of $A$ and $p$ for the data set where $L$ is > 0.40 pc and $T_k$ is < 250 K. Here the most probable values of $A$ and $p$ are 1.14 and 0.67 respectively.

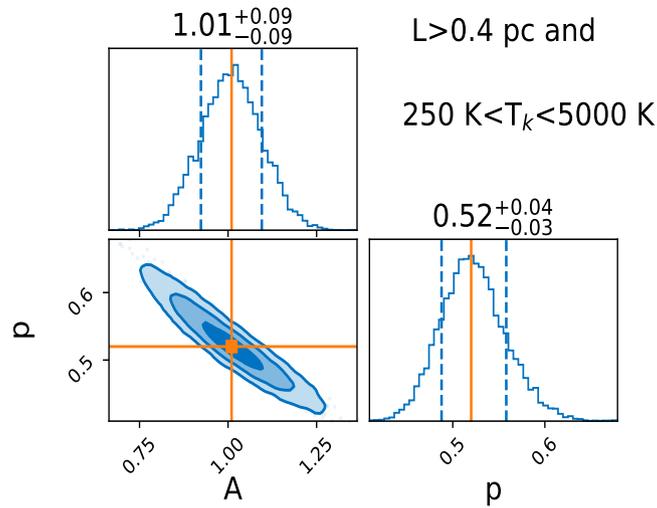

**Figure 10:** Histogram plot and the phase space diagram of $A$ and $p$ for the data set where $L$ is > 0.40 pc and and 250 K < $T_k$ < 5000 K. Here the most probable values of $A$ and $p$ are 1.01 and 0.52 respectively.

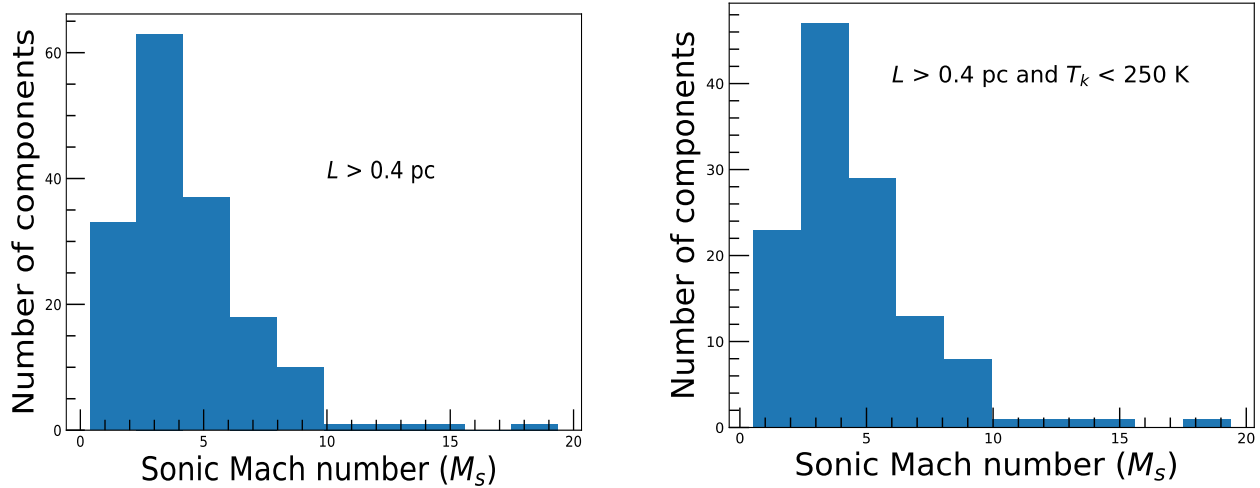

**Figure 11**: Left: Histogram plot of the sonic Mach number ($M_s$) of the gas components for which $L$ is > 0.4 pc. Right: Histogram plot of the sonic Mach number ($M_s$) of the gas components for which $L$ is > 0.4 pc and $T_k$ is < 250 K.

Table 1. : Col.1: Different dataset based on their length-scales ($L$). Col. 2: Median values and $1\sigma$ uncertainties (in parentheses) of Spearman correlation coefficients ($S$) for the same dataset.

| Data | Spearman correlation coefficient ($S$) |
|---|---|
| L >8.0 pc | 0.41 (0.13) |
| L >6.0 pc | 0.49 (0.11) |
| L >4.0 pc | 0.57 (0.08) |
| L >1.0 pc | 0.58 (0.05) |
| L >0.8 pc | 0.58 (0.04) |
| L >0.6 pc | 0.60 (0.04) |
| L >0.5 pc | 0.63 (0.04) |
| L >0.4 pc | 0.64 (0.04) |
| L >0.3 pc | 0.62 (0.03) |
| L >0.2 pc | 0.60 (0.03) |
| L >0.1 pc | 0.59 (0.03) |
| L >0.06 pc | 0.59 (0.03) |
| L >0.03 pc | 0.56 (0.02) |
| L >0.01 pc | 0.55 (0.02) |
| whole data | 0.54 (0.02) |
| L <0.4 pc | 0.13 (0.04) |

Table 2. : Col. 1: Source names of the background continuum objects toward which absorption spectra of H I are observed in different surveys. Col. 2 and Col. 3: Longitudes and latitudes of these background continuum sources respectively. Col. 4: Peak optical depths and their associated errors. Col. 5: Full width at half maxima of the components and their corresponding errors. Col. 6: Spin temperatures and their associated errors. Col. 7: Kinetic temperatures and their associated errors. Col. 8: Column densities of the gas cloud components. Col. 9: Non-thermal velocity dispersions and their associated errors. Col. 10: Length-scales and their associated errors. Col 11: References from where we have taken the data. $'a'$ denotes the work of Murray *et al.* (2018), $'b'$ denotes the work of Heiles & Troland (2003a), $'c'$ denotes the work of Stanimirović *et al.* (2014), and $'d'$ denotes the work of Patra *et al.* (2018).

| Source | Longitude | Latitude | $\tau_{peak}(\Delta\tau_{peak})$ | $V_{FWHM}(\Delta V_{FWHM})$ | $T_s(\Delta T_s)$ | $T_k(\Delta T_k)$ | $N_{HI}$ | $\sigma_{nth}(\Delta\sigma_{nth})$ | $L(\Delta L)$ | Ref. |
|--------|-----------|----------|------|------|------|------|------|------|------|------|
| (name) | (deg.) | (deg.) | | (km sec$^{-1}$) | (K) | (K) | ($\times 10^{20}$ cm$^{-2}$) | (km sec$^{-1}$) | (pc) | |
| J0022 | 107.462 | -61.748 | 0.018±0.001 | 2.8±0.1 | 83±2 | 83±2 | 0.08 | 0.85± 0.06 | 0.05±0.01 | *a* |
| J0022 | 107.462 | -61.748 | 0.008±0.001 | 10.3±0.3 | 567±19 | 605±20 | 0.88 | 3.76±0.15 | 4.38±1.12 | *a* |
| 3C018A | 118.623 | -52.732 | 0.565±0.007 | 2.5±0.1 | 17±1 | 17±1 | 0.48 | 0.99±0.05 | 0.07±0.02 | *a* |
| 3C018A | 118.623 | -52.732 | 0.134±0.003 | 5.5±0.2 | 196±5 | 208±5 | 2.81 | 1.93±0.10 | 4.80±1.06 | *a* |
| 3C018B | 118.616 | -52.719 | 0.524±0.006 | 2.4±0.1 | 16±2 | 16±2 | 0.41 | 0.95±0.05 | 0.05±0.02 | *a* |
| 3C018B | 118.616 | -52.719 | 0.149±0.005 | 6.2±0.1 | 162±4 | 171±4 | 2.89 | 2.35±0.05 | 4.07±0.89 | *a* |
| 3C041A | 131.379 | -29.075 | 0.033±0.004 | 8.9±0.2 | 351±7 | 374±8 | 1.99 | 3.35±0.10 | 6.12±1.51 | *a* |
| 3C041B | 131.374 | -29.070 | 0.022±0.005 | 1.7±0.2 | 14±3 | 14±3 | 0.01 | 0.64±0.10 | 0.001±0.0006 | *a* |
| 3C041B | 131.374 | -29.070 | 0.042±0.006 | 7.1±0.2 | 150±5 | 159±5 | 0.87 | 2.79±0.09 | 1.14±0.30 | *a* |
| 3C48 | 133.963 | -28.719 | 0.008±0.001 | 5.4±0.3 | 406±11 | 433±12 | 0.32 | 1.30±0.23 | 1.14±0.29 | *a* |
| 3C48 | 133.963 | -28.719 | 0.016±0.001 | 2.8±0.1 | 79±2 | 79±2 | 0.07 | 0.87±0.06 | 0.05±0.10 | *a* |
| 3C48 | 133.963 | -28.719 | 0.020±0.001 | 9.1±0.1 | 346±3 | 369±3 | 1.20 | 3.45±0.05 | 3.64±0.79 | *a* |
| 3C48 | 133.963 | -28.719 | 0.040±0.001 | 2.3±0.1 | 38±2 | 38±2 | 0.07 | 0.80±0.05 | 0.02±0.005 | *a* |
| 4C15.05 | 147.930 | -44.043 | 0.057±0.001 | 2.6±0.1 | 46±3 | 46±3 | 0.13 | 0.92±0.05 | 0.049±0.012 | *a* |
| 4C15.05 | 147.930 | -44.043 | 0.037±0.001 | 9.9±0.1 | 438±12 | 467±13 | 3.11 | 3.72±0.05 | 11.94±2.62 | *a* |
| 4C15.05 | 147.930 | -44.043 | 0.048±0.001 | 2.7±0.1 | 59±6 | 59±6 | 0.15 | 0.91±0.06 | 0.07±0.02 | *a* |
| 3C78 | 174.858 | -44.514 | 0.147±0.004 | 3.1±0.1 | 65±3 | 65±3 | 0.57 | 1.09±0.05 | 0.30±0.07 | *a* |
| 3C78 | 174.858 | -44.514 | 0.029±0.004 | 14.9±0.6 | 385±18 | 410±19 | 3.17 | 6.05±0.27 | 10.69±2.91 | *a* |
| 3C78 | 174.858 | -44.514 | 0.196±0.004 | 1.9±0.1 | 39±1 | 39±1 | 0.29 | 0.57±0.06 | 0.09±0.02 | *a* |
| 3C78 | 174.858 | -44.514 | 1.369±0.006 | 2.2±0.1 | 39±7 | 39±7 | 2.32 | 0.74±0.07 | 0.74±0.31 | *a* |
| 3C78 | 174.858 | -44.514 | 0.089±0.005 | 4.1±0.1 | 175±10 | 185±11 | 1.25 | 1.23±0.07 | 1.90±0.47 | *a* |
| 4C16.09 | 166.636 | -33.596 | 0.014±0.001 | 1.3±0.1 | 25±4 | 25±4 | 0.01 | 0.31±0.09 | 0.002±0.0008 | *a* |
| 4C16.09 | 166.636 | -33.596 | 0.436±0.002 | 4.2±0.1 | 106±3 | 112±3 | 3.74 | 1.50±0.05 | 3.44±0.76 | *a* |
| 4C16.09 | 166.636 | -33.596 | 0.227±0.004 | 1.8±0.1 | 54±10 | 54±10 | 0.44 | 0.37±0.14 | 0.20±0.08 | *a* |
| 4C16.09 | 166.636 | -33.596 | 0.086±0.001 | 5.4±0.1 | 260±3 | 276±3 | 2.37 | 1.72±0.06 | 5.38±1.15 | *a* |
| 3C111A | 161.676 | -8.820 | 0.013±0.003 | 5.7±1.0 | 487±44 | 519±47 | 0.69 | 1.25±0.83 | 2.95±1.18 | *a* |
| 3C111A | 161.676 | -8.820 | 0.041±0.003 | 2.6±0.1 | 131±6 | 139±6 | 0.27 | 0.27±0.20 | 0.31±0.08 | *a* |
| 3C111A | 161.676 | -8.820 | 0.322±0.004 | 2.6±0.1 | 34±5 | 34±5 | 0.56 | 0.97±0.05 | 0.16±0.06 | *a* |
| 3C111A | 161.676 | -8.820 | 0.172±0.005 | 8.6±0.3 | 248±10 | 263±11 | 7.18 | 3.34±0.14 | 15.55±3.58 | *a* |
| 3C111A | 161.676 | -8.820 | 0.790±0.009 | 3.9±0.1 | 77±4 | 77±4 | 5.57 | 1.45±0.05 | 2.89±0.68 | *a* |
| 3C111A | 161.676 | -8.820 | 0.643±0.020 | 3.3±0.1 | 93±8 | 96±11 | 3.85 | 1.08±0.07 | 3.05±0.89 | *a* |
| 3C111A | 161.676 | -8.820 | 0.232±0.004 | 5.2±0.1 | 129±1 | 136±1 | 3.02 | 1.94±0.05 | 3.39±0.72 | *a* |
| 3C111B | 161.686 | -8.788 | 0.019±0.002 | 1.7±0.1 | 26±1 | 26±1 | 0.02 | 0.55±0.06 | 0.004±0.001 | *a* |
| 3C111B | 161.686 | -8.788 | 0.111±0.003 | 2.7±0.1 | 30±5 | 30±5 | 0.17 | 1.03±0.05 | 0.04±0.02 | *a* |
| 3C111B | 161.686 | -8.788 | 0.092±0.004 | 12.1±0.3 | 409±10 | 436±11 | 8.86 | 4.77±0.14 | 32±7.05 | *a* |
| 3C111B | 161.686 | -8.788 | 0.400±0.005 | 2.8±0.1 | 14±8 | 14±8 | 0.32 | 1.14±0.05 | 0.04±0.04 | *a* |
| 3C111B | 161.686 | -8.788 | 0.967±0.020 | 3.7±0.1 | 56±5 | 56±5 | 3.88 | 1.42±0.05 | 1.79±0.50 | *a* |
| 3C111B | 161.686 | -8.788 | 0.347±0.054 | 2.7±0.1 | 58±13 | 58±13 | 1.07 | 0.91±0.08 | 0.51±0.27 | *a* |
| 3C111B | 161.686 | -8.788 | 0.342±0.032 | 5.6±0.4 | 120±12 | 127±13 | 4.51 | 2.15±0.19 | 4.70±1.48 | *a* |

| Source | Longitude | Latitude | $\tau_{peak}(\Delta\tau_{peak})$ | $V_{FWHM}(\Delta V_{FWHM})$ | $T_s(\Delta T_s)$ | $T_k(\Delta T_k)$ | $N_{HI}$ | $\sigma_{nth}(\Delta\sigma_{nth})$ | $L(\Delta L)$ | Ref. |
|---|---|---|---|---|---|---|---|---|---|---|
| (name) | (deg.) | (deg.) | | (km sec$^{-1}$) | (K) | (K) | ($\times 10^{20}$ cm$^{-2}$) | (km sec$^{-1}$) | (pc) | |
| 3C111C | 161.671 | -8.843 | 0.027±0.004 | 2.4±0.2 | 35±3 | 35±3 | 0.04 | 0.87±0.10 | 0.01±0.00 | a |
| 3C111C | 161.671 | -8.843 | 0.029±0.004 | 2.6±0.2 | 39±2 | 39±2 | 0.06 | 0.95±0.10 | 0.02±0.01 | a |
| 3C111C | 161.671 | -8.843 | 0.282±0.045 | 2.1±0.1 | 20±4 | 20±4 | 0.23 | 0.79±0.05 | 0.04±0.02 | a |
| 3C111C | 161.671 | -8.843 | 0.088±0.027 | 2.5±0.6 | 90±34 | 90±38 | 0.39 | 0.62±0.50 | 0.29±0.27 | a |
| 3C111C | 161.671 | -8.843 | 0.182±0.011 | 11.2±0.6 | 118±28 | 125±32 | 4.69 | 4.65±0.26 | 4.81±2.62 | a |
| 3C111C | 161.671 | -8.843 | 0.816±0.050 | 2.3±0.1 | 20±11 | 20±11 | 0.75 | 0.89±0.07 | 0.12±0.14 | a |
| 3C111C | 161.671 | -8.843 | 0.456±0.044 | 7.3±0.3 | 108±18 | 114±22 | 7.00 | 2.94±0.14 | 6.57±2.80 | a |
| 3C120 | 190.373 | -27.397 | 0.617±0.005 | 1.3±0.1 | 20±9 | 20±9 | 0.31 | 0.37±0.11 | 0.05±0.05 | a |
| 3C120 | 190.373 | -27.397 | 1.681±0.004 | 4.4±0.1 | 44±4 | 44±4 | 6.35 | 1.77±0.05 | 2.30±0.64 | a |
| 3C120 | 190.373 | -27.397 | 0.776±0.012 | 2.3±0.1 | 23±5 | 23±5 | 0.82 | 0.87±0.05 | 0.16±0.08 | a |
| 3C123A | 170.584 | -11.660 | 0.064±0.001 | 3.1±0.1 | 26±2 | 26±2 | 0.10 | 1.23±0.05 | 0.02±0.01 | a |
| 3C123A | 170.584 | -11.660 | 0.379±0.035 | 2.2±0.1 | 19±14 | 19±14 | 0.32 | 0.85±0.08 | 0.05±0.07 | a |
| 3C123A | 170.584 | -11.660 | 0.044±0.004 | 10.1±0.3 | 619±79 | 660±84 | 5.36 | 3.60±0.18 | 29.10±10.03 | a |
| 3C123A | 170.584 | -11.660 | 0.810±0.071 | 4.4±0.1 | 67±25 | 67±26 | 4.67 | 1.71±0.08 | 2.57±2.05 | a |
| 3C123A | 170.584 | -11.660 | 0.628±0.037 | 1.8±0.1 | 18±11 | 18±11 | 0.40 | 0.66±0.08 | 0.06±0.07 | a |
| 3C123A | 170.584 | -11.660 | 0.008±0.001 | 4.0±0.3 | 124±9 | 131±10 | 0.08 | 1.34±0.16 | 0.09±0.03 | a |
| 3C123B | 170.578 | -11.659 | 0.064±0.001 | 3.2±0.1 | 16±2 | 16±2 | 0.07 | 1.31±0.04 | 0.01±0.00 | a |
| 3C123B | 170.578 | -11.659 | 1.648±0.005 | 4.3±0.1 | 36±14 | 36±14 | 5.08 | 1.74±0.06 | 1.50±1.21 | a |
| 3C123B | 170.578 | -11.659 | 0.042±0.003 | 10.9±0.3 | 664±52 | 708±56 | 5.98 | 3.95±0.16 | 34.83±9.52 | a |
| 3C132 | 178.862 | -12.522 | 0.197±0.009 | 7.2±0.1 | 165±12 | 175±13 | 4.57 | 2.81±0.05 | 6.56±1.71 | a |
| 3C132 | 178.862 | -12.522 | 1.472±0.005 | 2.0±0.1 | 27±19 | 27±19 | 1.55 | 0.71±0.12 | 0.34±0.49 | a |
| 3C132 | 178.862 | -12.522 | 0.230±0.003 | 6.7±0.1 | 80±4 | 80±4 | 2.39 | 2.73±0.04 | 1.57±0.37 | a |
| 3C133 | 177.725 | -9.913 | 0.033±0.005 | 9.5±0.3 | 537±28 | 573±30 | 3.26 | 3.40±0.16 | 15.35±4.32 | a |
| 3C133 | 177.725 | -9.913 | 0.134±0.007 | 2.4±0.1 | 26±16 | 26±16 | 0.17 | 0.91±0.09 | 0.04±0.05 | a |
| 3C133 | 177.725 | -9.913 | 0.240±0.011 | 9.6±0.1 | 283±13 | 301±14 | 12.67 | 3.76±0.05 | 31.36±7.36 | a |
| 3C133 | 177.725 | -9.913 | 0.706±0.010 | 2.7±0.1 | 11±15 | 11±15 | 0.44 | 1.11±0.07 | 0.04±0.11 | a |
| 3C133 | 177.725 | -9.913 | 0.845±0.029 | 1.5±0.1 | 23±19 | 23±19 | 0.58 | 0.46±0.18 | 0.11±0.18 | a |
| 3C133 | 177.725 | -9.913 | 0.780±0.022 | 2.6±0.1 | 46±9 | 46±9 | 1.82 | 0.92±0.07 | 0.69±0.31 | a |
| 3C138 | 187.405 | -11.343 | 0.033±0.001 | 2.3±0.1 | 60±2 | 60±2 | 0.09 | 0.68±0.06 | 0.04±0.01 | a |
| 3C138 | 187.405 | -11.343 | 0.026±0.002 | 4.3±0.2 | 297±11 | 316±12 | 0.65 | 0.85±0.19 | 1.69±0.41 | a |
| 3C138 | 187.405 | -11.343 | 0.322±0.003 | 4.3±0.1 | 125±3 | 132±3 | 3.39 | 1.50±0.05 | 3.68±0.80 | a |
| 3C138 | 187.405 | -11.343 | 0.003±0.003 | 37.4±4.9 | 1427±302 | 1640±405 | 3.52 | 15.45±2.14 | 47.48±53.55 | a |
| 3C138 | 187.405 | -11.343 | 1.057±0.005 | 2.4±0.1 | 39±7 | 39±7 | 1.92 | 0.85±0.06 | 0.62±0.26 | a |
| 3C138 | 187.405 | -11.343 | 0.432±0.003 | 3.0±0.1 | 66±1 | 66±1 | 1.68 | 1.04±0.05 | 0.91±0.20 | a |
| PKS0531 | 186.762 | -7.108 | 0.001±0.001 | 12.1±1.5 | 2049±400 | 2725±816 | 0.68 | 1.97±2.39 | 15.24±17.41 | a |
| PKS0531 | 186.762 | -7.108 | 0.007±0.001 | 3.5±0.2 | 24±4 | 24±4 | 0.01 | 1.42±0.09 | 0.002±0.001 | a |
| PKS0531 | 186.762 | -7.108 | 0.006±0.001 | 6.2±0.3 | 240±130 | 255±139 | 0.16 | 2.20±0.30 | 0.34±0.38 | a |
| PKS0531 | 186.762 | -7.108 | 0.403±0.009 | 1.9±0.1 | 28±26 | 28±26 | 0.44 | 0.65±0.17 | 0.10±0.19 | a |
| PKS0531 | 186.762 | -7.108 | 0.141±0.005 | 9.7±0.1 | 542±25 | 578±27 | 14.48 | 3.50±0.06 | 68.83±16.03 | a |
| PKS0531 | 186.762 | -7.108 | 0.190±0.003 | 2.4±0.1 | 100±10 | 106±13 | 0.88 | 0.41±0.17 | 0.76±0.24 | a |
| 3C147 | 161.686 | 10.298 | 0.013±0.001 | 2.0±0.1 | 65±5 | 65±5 | 0.03 | 0.43±0.10 | 0.016±0.004 | a |
| 3C147 | 161.686 | 10.298 | 0.054±0.009 | 1.9±0.1 | 36±8 | 36±8 | 0.07 | 0.59±0.08 | 0.02±0.01 | a |
| 3C147 | 161.686 | 10.298 | 0.120±0.021 | 6.4±0.3 | 218±37 | 231±40 | 3.25 | 2.34±0.16 | 6.18±2.72 | a |

| Source | Longitude | Latitude | $\tau_{peak}(\Delta\tau_{peak})$ | $V_{FWHM}(\Delta V_{FWHM})$ | $T_s(\Delta T_s)$ | $T_k(\Delta T_k)$ | $N_{HI}$ | $\sigma_{nth}(\Delta\sigma_{nth})$ | $L(\Delta L)$ | Ref. |
|---|---|---|---|---|---|---|---|---|---|---|
| (name) | (deg.) | (deg.) | | (km sec$^{-1}$) | (K) | (K) | ($\times 10^{20}$ cm$^{-2}$) | (km sec$^{-1}$) | (pc) | |
| 3C147 | 161.686 | 10.298 | 0.165±0.020 | 2.3±0.1 | 37±6 | 37±6 | 0.27 | 0.81±0.06 | 0.08±0.03 | a |
| 3C147 | 161.686 | 10.298 | 0.278±0.004 | 5.9±0.1 | 109±9 | 115±10 | 3.49 | 2.31±0.05 | 3.31±0.89 | a |
| 3C147 | 161.686 | 10.298 | 0.035±0.001 | 1.6±0.1 | 25±3 | 25±3 | 0.03 | 0.51±0.06 | 0.01±0.00 | a |
| 3C154 | 185.592 | 4.003 | 0.061±0.005 | 3.1±0.2 | 25±3 | 25±3 | 0.10 | 1.24±0.09 | 0.02±0.01 | a |
| 3C154 | 185.592 | 4.003 | 0.012±0.005 | 7.4±1.4 | 587±240 | 626±256 | 1.05 | 2.17±0.99 | 5.41±5.20 | a |
| 3C154 | 185.592 | 4.003 | 0.123±0.020 | 14.0±0.5 | 762±130 | 813±139 | 25.55 | 5.35±0.26 | 170.83±74.18 | a |
| 3C154 | 185.592 | 4.003 | 1.300±0.015 | 2.3±0.1 | 10±15 | 10±15 | 0.62 | 0.93±0.08 | 0.05±0.15 | a |
| 3C154 | 185.592 | 4.003 | 0.733±0.016 | 3.9±0.1 | 19±17 | 19±17 | 1.11 | 1.61±0.06 | 0.17±0.31 | a |
| 3C154 | 185.592 | 4.003 | 0.521±0.004 | 2.2±0.1 | 19±3 | 19±3 | 0.44 | 0.85±0.05 | 0.07±0.03 | a |
| PKS0742 | 209.797 | 16.592 | 0.009±0.001 | 3.2±0.1 | 164±9 | 174±10 | 0.09 | 0.64±0.11 | 0.13±0.03 | a |
| 3C225A | 220.010 | 44.008 | 0.043±0.002 | 1.8±0.1 | 22±0 | 22±0 | 0.03 | 0.63±0.05 | 0.01±0.00 | a |
| 3C225A | 220.010 | 44.008 | 0.020±0.002 | 4.8±0.3 | 60±1 | 60±1 | 0.11 | 1.91±0.14 | 0.05±0.01 | a |
| 3C225A | 220.010 | 44.008 | 0.048±0.002 | 2.5±0.1 | 22±0 | 22±0 | 0.05 | 0.97±0.05 | 0.01±0.00 | a |
| 3C225A | 220.010 | 44.008 | 0.013±0.002 | 7.7±0.3 | 327±11 | 348±12 | 0.64 | 2.80±0.15 | 1.83±0.50 | a |
| 3C225A | 220.010 | 44.008 | 0.805±0.002 | 1.3±0.1 | 11±2 | 11±2 | 0.23 | 0.46±0.05 | 0.02±0.01 | a |
| 3C225B | 220.011 | 44.009 | 0.044±0.003 | 2.0±0.1 | 18±0 | 18±0 | 0.03 | 0.76±0.05 | 0.00±0.00 | a |
| 3C225B | 220.011 | 44.009 | 0.023±0.003 | 4.0±0.3 | 145±4 | 153±4 | 0.26 | 1.27±0.17 | 0.33±0.09 | a |
| 3C225B | 220.011 | 44.009 | 0.053±0.003 | 2.4±0.1 | 20±0 | 20±0 | 0.05 | 0.93±0.05 | 0.01±0.00 | a |
| 3C225B | 220.011 | 44.009 | 0.013±0.003 | 8.3±0.4 | 458±17 | 488±18 | 1.00 | 2.90±0.20 | 4.02±1.30 | a |
| 3C225B | 220.011 | 44.009 | 0.774±0.003 | 1.3±0.1 | 14±0 | 14±0 | 0.28 | 0.43±0.05 | 0.03±0.01 | a |
| 3C237 | 232.117 | 46.627 | 0.006±0.001 | 15.0±0.6 | 382±17 | 407±18 | 0.63 | 6.10±0.27 | 2.11±0.60 | a |
| 3C237 | 232.117 | 46.627 | 0.415±0.001 | 1.2±0.1 | 13±0 | 13±0 | 0.13 | 0.39±0.06 | 0.01±0.00 | a |
| 3C245A | 233.124 | 56.300 | 0.010±0.002 | 5.3±0.3 | 385±30 | 410±32 | 0.40 | 1.29±0.24 | 1.35±0.45 | a |
| 1055+018 | 251.511 | 52.774 | 0.006±0.001 | 7.1±0.3 | 941±30 | 1005±36 | 0.83 | 0.89±0.46 | 6.86±1.92 | a |
| 3C263.1 | 227.201 | 73.766 | 0.020±0.001 | 2.0±0.1 | 35±0 | 35±0 | 0.03 | 0.66±0.05 | 0.01±0.00 | a |
| 3C273 | 289.945 | 64.359 | 0.019±0.001 | 2.3±0.1 | 17±0 | 17±0 | 0.02 | 0.90±0.05 | 0.003±0.001 | a |
| 3C273 | 289.945 | 64.359 | 0.005±0.001 | 6.4±0.3 | 455±47 | 485±50 | 0.29 | 1.84±0.22 | 1.16±0.42 | a |
| 4C32.44 | 67.234 | 81.048 | 0.018±0.001 | 2.8±0.1 | 112±2 | 118±2 | 0.11 | 0.66±0.08 | 0.11±0.02 | a |
| 4C32.44 | 67.234 | 81.048 | 0.004±0.001 | 3.7±0.3 | 255±17 | 271±18 | 0.07 | 0.48±0.45 | 0.16±0.06 | a |
| 3C286 | 56.524 | 80.675 | 0.006±0.001 | 2.4±0.1 | 76±2 | 76±2 | 0.02 | 0.64±0.07 | 0.01±0.00 | a |
| 3C286 | 56.524 | 80.675 | 0.005±0.001 | 3.2±0.2 | 60±2 | 60±2 | 0.02 | 1.16±0.10 | 0.01±0.00 | a |
| 3C286 | 56.524 | 80.675 | 0.007±0.001 | 4.3±0.1 | 78±2 | 78±2 | 0.05 | 1.64±0.05 | 0.03±0.01 | a |
| 4C12.50 | 347.223 | 70.172 | 0.016±0.001 | 6.6±0.1 | 318±11 | 339±12 | 0.66 | 2.25±0.06 | 1.84±0.42 | a |
| 4C12.50 | 347.223 | 70.172 | 0.077±0.002 | 2.2±0.1 | 97±1 | 102±1 | 0.33 | 0.16±0.25 | 0.28±0.06 | a |
| 3C298 | 352.160 | 60.666 | 0.019±0.001 | 3.6±0.1 | 112±10 | 118±11 | 0.15 | 1.17±0.07 | 0.15±0.04 | a |
| UGC09799 | 9.417 | 50.120 | 0.058±0.010 | 2.7±0.1 | 58±4 | 58±4 | 0.17 | 0.91±0.06 | 0.08±0.02 | a |
| 4C04.51 | 7.292 | 47.747 | 0.002±0.001 | 14.0±1.5 | 114±8 | 120±8 | 0.07 | 5.86±0.65 | 0.07±0.04 | a |
| 3C327.1A | 12.181 | 37.006 | 0.126±0.006 | 3.4±0.1 | 140±3 | 148±3 | 1.16 | 0.93±0.07 | 1.41±0.31 | a |
| 3C327.1A | 12.181 | 37.006 | 0.425±0.010 | 1.9±0.1 | 63±4 | 63±4 | 1.00 | 0.36±0.11 | 0.52±0.13 | a |
| 3C327.1A | 12.181 | 37.006 | 0.401±0.008 | 2.2±0.1 | 69±3 | 69±3 | 1.18 | 0.55±0.08 | 0.67±0.16 | a |
| 3C327.1B | 12.182 | 37.003 | 0.118±0.006 | 3.2±0.1 | 142±3 | 150±3 | 1.05 | 0.78±0.08 | 1.30±0.29 | a |
| 3C327.1B | 12.182 | 37.003 | 0.359±0.011 | 1.9±0.1 | 74±4 | 74±4 | 1.01 | 0.20±0.19 | 0.61±0.15 | a |
| 3C327.1B | 12.182 | 37.003 | 0.419±0.008 | 2.2±0.1 | 68±3 | 68±3 | 1.20 | 0.56±0.07 | 0.67±0.16 | a |

| Source | Longitude | Latitude | $\tau_{peak}(\Delta\tau_{peak})$ | $V_{FWHM}(\Delta V_{FWHM})$ | $T_s(\Delta T_s)$ | $T_k(\Delta T_k)$ | $N_{HI}$ | $\sigma_{nth}(\Delta\sigma_{nth})$ | $L(\Delta L)$ | Ref. |
|---|---|---|---|---|---|---|---|---|---|---|
| (name) | (deg.) | (deg.) | | (km sec$^{-1}$) | (K) | (K) | ($\times 10^{20}$ cm$^{-2}$) | (km sec$^{-1}$) | (pc) | |
| PKS1607 | 44.171 | 46.203 | 0.128±0.001 | 2.0±0.1 | 26±1 | 26±1 | 0.13 | 0.71±0.05 | 0.03±0.01 | a |
| PKS1607 | 44.171 | 46.203 | 0.013±0.001 | 6.5±0.3 | 114±10 | 120±11 | 0.19 | 2.57±0.14 | 0.19±0.05 | a |
| PKS1607 | 44.171 | 46.203 | 0.064±0.004 | 3.9±0.1 | 115±10 | 122±11 | 0.56 | 1.32±0.06 | 0.56±0.16 | a |
| 3C346 | 35.332 | 35.769 | 0.035±0.003 | 4.5±0.1 | 133±3 | 141±3 | 0.40 | 1.58±0.05 | 0.46±0.11 | a |
| 3C346 | 35.332 | 35.769 | 0.279±0.004 | 2.0±0.1 | 58±4 | 58±4 | 0.62 | 0.49±0.08 | 0.30±0.08 | a |
| 3C346 | 35.332 | 35.769 | 0.197±0.004 | 1.9±0.1 | 59±4 | 59±4 | 0.43 | 0.40±0.09 | 0.21±0.05 | a |
| 3C390 | 41.112 | 5.773 | 0.146±0.003 | 3.7±0.2 | 221±7 | 234±8 | 2.31 | 0.73±0.19 | 4.45±1.01 | a |
| 3C390 | 41.112 | 5.773 | 0.082±0.003 | 5.7±0.2 | 331±7 | 352±8 | 3.01 | 1.72±0.12 | 8.72±1.93 | a |
| 3C390 | 41.112 | 5.773 | 0.043±0.003 | 2.0±0.1 | 27±4 | 27±4 | 0.05 | 0.71±0.06 | 0.01±0.00 | a |
| 3C390 | 41.112 | 5.773 | 0.095±0.003 | 9.8±0.1 | 415±6 | 442±6 | 7.51 | 3.70±0.05 | 27.32±5.88 | a |
| 3C390 | 41.112 | 5.773 | 0.067±0.003 | 1.4±0.1 | 30±4 | 30±4 | 0.06 | 0.32±0.09 | 0.01±0.01 | a |
| 4C33.48 | 66.389 | 8.371 | 0.170±0.006 | 2.2±0.1 | 26±5 | 26±5 | 0.20 | 0.81±0.06 | 0.04±0.02 | a |
| 4C33.48 | 66.389 | 8.371 | 0.139±0.006 | 6.8±0.1 | 75±13 | 75±13 | 1.39 | 2.78±0.05 | 0.86±0.35 | a |
| 3C409A | 63.398 | -6.121 | 0.443±0.004 | 3.2±0.1 | 64±9 | 64±9 | 1.77 | 1.15±0.06 | 0.93±0.33 | a |
| 3C409A | 63.398 | -6.121 | 0.332±0.005 | 3.0±0.1 | 122±11 | 129±12 | 2.36 | 0.75±0.10 | 2.50±0.70 | a |
| 3C409A | 63.398 | -6.121 | 0.732±0.007 | 2.1±0.1 | 13±5 | 13±5 | 0.41 | 0.83±0.05 | 0.04±0.04 | a |
| 3C409A | 63.398 | -6.121 | 0.440±0.008 | 6.5±0.1 | 145±20 | 153±21 | 8.11 | 2.52±0.06 | 10.23±3.56 | a |
| 3C409A | 63.398 | -6.121 | 0.735±0.006 | 1.7±0.1 | 12±18 | 12±18 | 0.29 | 0.65±0.12 | 0.03±0.09 | a |
| 3C409A | 63.398 | -6.121 | 0.020±0.003 | 4.3±0.2 | 366±39 | 390±42 | 0.63 | 0.33±0.70 | 2.02±0.68 | a |
| 3C409B | 63.398 | -6.122 | 0.429±0.003 | 2.9±0.1 | 49±9 | 49±9 | 1.18 | 1.05±0.06 | 0.48±0.20 | a |
| 3C409B | 63.398 | -6.122 | 0.280±0.007 | 3.0±0.1 | 120±13 | 127±14 | 1.97 | 0.76±0.10 | 2.06±0.63 | a |
| 3C409B | 63.398 | -6.122 | 0.631±0.081 | 1.9±0.1 | 66±22 | 66±22 | 1.51 | 0.32±0.30 | 0.82±0.58 | a |
| 3C409B | 63.398 | -6.122 | 0.106±0.004 | 13.1±0.2 | 285±12 | 303±13 | 7.67 | 5.33±0.09 | 19.12±4.41 | a |
| 3C409B | 63.398 | -6.122 | 0.890±0.065 | 3.0±0.1 | 47±13 | 47±13 | 2.45 | 1.11±0.07 | 0.95±0.57 | a |
| 3C410A | 69.212 | -3.769 | 0.014±0.002 | 1.5±0.1 | 15±3 | 15±3 | 0.01 | 0.53±0.06 | 0.001±0.001 | a |
| 3C410A | 69.212 | -3.769 | 0.613±0.018 | 3.6±0.1 | 48±12 | 48±12 | 2.07 | 1.39±0.06 | 0.82±0.44 | a |
| 3C410A | 69.212 | -3.769 | 0.648±0.023 | 3.2±0.1 | 49±14 | 49±14 | 2.02 | 1.20±0.07 | 0.81±0.50 | a |
| 3C410A | 69.212 | -3.769 | 1.693±0.150 | 3.2±0.1 | 70±13 | 70±13 | 7.33 | 1.13±0.07 | 4.22±1.84 | a |
| 3C410A | 69.212 | -3.769 | 0.575±0.096 | 4.4±0.4 | 111±21 | 117±25 | 5.46 | 1.59±0.21 | 5.27±2.58 | a |
| 3C410A | 69.212 | -3.769 | 0.186±0.003 | 3.3±0.1 | 71±8 | 71±8 | 0.86 | 1.17±0.06 | 0.50±0.16 | a |
| 3C410A | 69.212 | -3.769 | 0.060±0.003 | 5.2±0.1 | 69±14 | 69±14 | 0.43 | 2.08±0.05 | 0.24±0.11 | a |
| 3C410B | 69.211 | -3.770 | 0.021±0.002 | 5.3±0.4 | 39±3 | 39±3 | 0.08 | 2.18±0.18 | 0.03±0.01 | a |
| 3C410B | 69.211 | -3.770 | 0.019±0.002 | 1.8±0.3 | 18±3 | 18±3 | 0.01 | 0.66±0.15 | 0.001±0.001 | a |
| 3C410B | 69.211 | -3.770 | 0.476±0.007 | 2.7±0.1 | 28±2 | 28±2 | 0.71 | 1.04±0.05 | 0.16±0.04 | a |
| 3C410B | 69.211 | -3.770 | 0.344±0.023 | 11.4±0.5 | 153±12 | 162±13 | 11.71 | 4.70±0.22 | 15.59±4.28 | a |
| 3C410B | 69.211 | -3.770 | 0.292±0.011 | 1.5±0.1 | 31±3 | 31±3 | 0.26 | 0.39±0.08 | 0.07±0.02 | a |
| 3C410B | 69.211 | -3.770 | 2.798±0.014 | 2.2±0.1 | 18±1 | 18±1 | 2.15 | 0.85±0.05 | 0.32±0.08 | a |
| 3C410B | 69.211 | -3.770 | 0.430±0.137 | 1.9±0.1 | 65±21 | 65±21 | 1.06 | 0.34±0.28 | 0.57±0.43 | a |
| 3C410B | 69.211 | -3.770 | 0.041±0.017 | 19.2±2.0 | 1006±403 | 1090±487 | 15.52 | 7.58±0.95 | 139.13±135.24 | a |
| 3C410B | 69.211 | -3.770 | 0.112±0.004 | 2.5±0.1 | 32±3 | 32±3 | 0.18 | 0.93±0.05 | 0.05±0.01 | a |
| B2050 | 78.858 | -5.124 | 0.007±0.001 | 3.4±0.2 | 141±9 | 149±10 | 0.06 | 0.92±0.14 | 0.07±0.02 | a |
| B2050 | 78.858 | -5.124 | 0.011±0.001 | 2.4±0.1 | 78±12 | 78±12 | 0.04 | 0.63±0.10 | 0.03±0.01 | a |
| B2050 | 78.858 | -5.124 | 0.043±0.002 | 10.0±0.3 | 453±29 | 483±31 | 3.79 | 3.75±0.15 | 15.05±3.80 | a |

| Source | Longitude | Latitude | $\tau_{peak}(\Delta\tau_{peak})$ | $V_{FWHM}(\Delta V_{FWHM})$ | $T_s(\Delta T_s)$ | $T_k(\Delta T_k)$ | $N_{HI}$ | $\sigma_{nth}(\Delta\sigma_{nth})$ | $L(\Delta L)$ | Ref. |
|---|---|---|---|---|---|---|---|---|---|---|
| (name) | (deg.) | (deg.) | | (km sec$^{-1}$) | (K) | (K) | ($\times 10^{20}$ cm$^{-2}$) | (km sec$^{-1}$) | (pc) | |
| B2050 | 78.858 | -5.124 | 0.039±0.002 | 2.0±0.1 | 13±5 | 13±5 | 0.02 | 0.78±0.05 | 0.002±0.002 | a |
| B2050 | 78.858 | -5.124 | 0.150±0.002 | 1.7±0.1 | 10±6 | 10±6 | 0.05 | 0.66±0.06 | 0.004±0.005 | a |
| B2050 | 78.858 | -5.124 | 0.074±0.002 | 3.8±0.1 | 28±6 | 28±6 | 0.15 | 1.54±0.05 | 0.03±0.02 | a |
| 3C433 | 74.475 | -17.697 | 0.181±0.010 | 1.5±0.1 | 21±9 | 21±9 | 0.11 | 0.48±0.10 | 0.02±0.02 | a |
| 3C433 | 74.475 | -17.697 | 0.304±0.009 | 3.9±0.1 | 78±5 | 78±5 | 1.82 | 1.45±0.05 | 1.17±0.29 | a |
| 3C433 | 74.475 | -17.697 | 0.080±0.005 | 2.1±0.1 | 29±5 | 29±5 | 0.09 | 0.75±0.06 | 0.02±0.01 | a |
| 3C433 | 74.475 | -17.697 | 0.059±0.004 | 2.5±0.1 | 29±2 | 29±2 | 0.08 | 0.94±0.05 | 0.02±0.01 | a |
| PKS2127 | 58.652 | -31.815 | 0.073±0.001 | 2.7±0.1 | 105±4 | 111±4 | 0.40 | 0.63±0.08 | 0.36±0.08 | a |
| PKS2127 | 58.652 | -31.815 | 0.013±0.001 | 8.3±0.2 | 769±43 | 820±46 | 1.59 | 2.38±0.15 | 10.73±2.70 | a |
| PKS2127 | 58.652 | -31.815 | 0.107±0.001 | 2.1±0.1 | 61±5 | 61±5 | 0.27 | 0.54±0.08 | 0.14±0.04 | a |
| J2136 | 55.473 | -35.578 | 0.096±0.002 | 2.2±0.1 | 28±3 | 28±3 | 0.12 | 0.80±0.05 | 0.03±0.01 | a |
| J2136 | 55.473 | -35.578 | 0.062±0.002 | 7.0±0.1 | 22±13 | 22±13 | 0.19 | 2.94±0.05 | 0.03±0.04 | a |
| J2136 | 55.473 | -35.578 | 0.079±0.002 | 2.5±0.1 | 98±4 | 103±5 | 0.37 | 0.52±0.09 | 0.31±0.07 | a |
| J2232 | 77.438 | -38.582 | 0.066±0.001 | 2.3±0.1 | 65±0 | 65±0 | 0.19 | 0.65±0.06 | 0.10±0.02 | a |
| J2232 | 77.438 | -38.582 | 0.056±0.002 | 2.0±0.1 | 58±3 | 58±3 | 0.13 | 0.49±0.08 | 0.06±0.02 | a |
| J2232 | 77.438 | -38.582 | 0.064±0.002 | 5.7±0.1 | 105±4 | 111±4 | 0.75 | 2.22±0.06 | 0.68±0.16 | a |
| J2232 | 77.438 | -38.582 | 0.099±0.002 | 2.0±0.1 | 32±4 | 32±4 | 0.12 | 0.68±0.06 | 0.03±0.01 | a |
| J2232 | 77.438 | -38.582 | 0.046±0.002 | 3.3±0.1 | 96±3 | 101±4 | 0.28 | 1.06±0.06 | 0.23±0.05 | a |
| 3C454.3 | 86.112 | -38.185 | 0.313±0.024 | 2.5±0.1 | 20±2 | 20±2 | 0.31 | 0.98±0.05 | 0.05±0.02 | a |
| 3C454.3 | 86.112 | -38.185 | 0.025±0.013 | 3.7±1.2 | 159±90 | 168±98 | 0.28 | 1.04±0.86 | 0.39±0.51 | a |
| 3C454.3 | 86.112 | -38.185 | 0.096±0.002 | 3.6±0.1 | 142±2 | 150±2 | 0.95 | 1.05±0.06 | 1.17±0.25 | a |
| 3C454.3 | 86.112 | -38.185 | 0.081±0.002 | 1.8±0.1 | 61±3 | 61±3 | 0.17 | 0.28±0.12 | 0.09±0.02 | a |
| 3C454.3 | 86.112 | -38.185 | 0.026±0.002 | 4.7±0.3 | 109±3 | 115±3 | 0.25 | 1.74±0.15 | 0.24±0.06 | a |
| 3C459 | 83.040 | -51.285 | 0.010±0.001 | 2.8±0.1 | 72±2 | 72±2 | 0.04 | 0.90±0.06 | 0.02±0.01 | a |
| 3C459 | 83.040 | -51.285 | 0.039±0.002 | 5.2±0.3 | 384±22 | 409±24 | 1.48 | 1.22±0.24 | 4.98±1.25 | a |
| 3C459 | 83.040 | -51.285 | 0.057±0.002 | 1.9±0.1 | 37±8 | 37±8 | 0.08 | 0.59±0.08 | 0.02±0.01 | a |
| 3C459 | 83.040 | -51.285 | 0.039±0.003 | 7.7±0.6 | 478±35 | 510±37 | 2.79 | 2.55±0.33 | 11.69±3.26 | a |
| 3C459 | 83.040 | -51.285 | 0.088±0.003 | 2.2±0.1 | 46±6 | 46±6 | 0.18 | 0.70±0.10 | 0.07±0.02 | a |
| 3C33-1 | 129.44 | -49.34 | 0.034±0.001 | 9.44±0.30 | 310±5 | 330±5 | 1.95 | 3.65±0.14 | 5.28±1.15 | b |
| 3C33-2 | 129.46 | -49.28 | 0.059±0.002 | 9.31±0.42 | 178±5 | 188±5 | 1.90 | 3.75±0.19 | 2.94±0.66 | b |
| 3C33 | 129.45 | -49.32 | 0.025±0.000 | 8.91±0.10 | 379±3 | 404±3 | 1.64 | 3.31±0.05 | 5.45±1.15 | b |
| 3C64 | 157.77 | -48.20 | 0.290±0.007 | 5.10±0.13 | 90±4 | 90±6 | 2.58 | 1.99±0.06 | 1.90±0.46 | b |
| 3C64 | 157.77 | -48.20 | 0.086±0.007 | 4.37±0.38 | 45±5 | 45±5 | 0.33 | 1.75±0.17 | 0.12±0.04 | b |
| 3C75-1 | 170.22 | -44.91 | 0.729±0.014 | 2.06±0.03 | 35±5 | 35±5 | 1.02 | 0.69±0.03 | 0.29±0.10 | b |
| 3C75-1 | 170.22 | -44.91 | 0.082±0.003 | 3.01±0.30 | 16±5 | 16±5 | 0.08 | 1.22±0.13 | 0.01±0.01 | b |
| 3C75-1 | 170.22 | -44.91 | 0.127±0.006 | 4.58±0.22 | 84±8 | 84±9 | 0.95 | 1.76±0.10 | 0.66±0.19 | b |
| 3C75-2 | 170.30 | -44.92 | 0.647±0.015 | 2.33±0.05 | 36±5 | 36±5 | 1.06 | 0.82±0.03 | 0.32±0.10 | b |
| 3C75-2 | 170.30 | -44.92 | 0.094±0.009 | 2.34±0.26 | 34±4 | 34±4 | 0.15 | 0.84±0.13 | 0.04±0.02 | b |
| 3C75-2 | 170.30 | -44.92 | 0.139±0.007 | 4.41±0.23 | 78±8 | 78±8 | 0.93 | 1.69±0.11 | 0.60±0.18 | b |
| 3C75 | 170.26 | -44.91 | 0.682±0.006 | 2.32±0.02 | 36±4 | 36±4 | 1.12 | 0.82±0.02 | 0.33±0.10 | b |
| 3C75 | 170.26 | -44.91 | 0.095±0.004 | 2.74±0.12 | 37±3 | 37±3 | 0.19 | 1.02±0.06 | 0.06±0.02 | b |
| 3C75 | 170.26 | -44.91 | 0.113±0.003 | 5.19±0.13 | 104±7 | 109±7 | 1.18 | 1.99±0.06 | 1.06±0.27 | b |
| 3C79 | 164.15 | -34.46 | 0.342±0.006 | 2.06±0.04 | 9±5 | 9±5 | 0.12 | 0.83±0.03 | 0.01±0.01 | b |

| Source | Longitude | Latitude | $\tau_{peak}(\Delta\tau_{peak})$ | $V_{FWHM}(\Delta V_{FWHM})$ | $T_s(\Delta T_s)$ | $T_k(\Delta T_k)$ | $N_{HI}$ | $\sigma_{nth}(\Delta\sigma_{nth})$ | $L(\Delta L)$ | Ref. |
|---|---|---|---|---|---|---|---|---|---|---|
| 3C79 | 164.15 | -34.46 | 0.082±0.007 | 1.18±0.12 | 20±8 | 20±8 | 0.04 | 0.30±0.14 | 0.01±0.01 | *b* |
| 3C79 | 164.15 | -34.46 | 0.139±0.005 | 2.85±0.14 | 53±8 | 53±8 | 0.41 | 1.01±0.08 | 0.18±0.07 | *b* |
| 3C79 | 164.15 | -34.46 | 0.148±0.004 | 14.09±0.27 | 156±18 | 167±19 | 6.30 | 5.87±0.12 | 8.53±2.71 | *b* |
| 3C93.1 | 160.04 | -15.91 | 0.860±0.080 | 2.99±0.24 | 35±13 | 35±13 | 1.73 | 1.15±0.12 | 0.49±0.39 | *b* |
| 3C93.1 | 160.04 | -15.91 | 1.243±0.117 | 2.54±0.13 | 29±11 | 29±11 | 1.77 | 0.96±0.08 | 0.42±0.34 | *b* |
| 3C98-1 | 179.86 | -31.09 | 0.081±0.004 | 3.23±0.18 | 21±6 | 21±6 | 0.11 | 1.31±0.08 | 0.02±0.01 | *b* |
| 3C98-1 | 179.86 | -31.09 | 0.209±0.011 | 1.47±0.09 | 41±17 | 41±17 | 0.25 | 0.22±0.33 | 0.08±0.07 | *b* |
| 3C98-1 | 179.86 | -31.09 | 0.368±0.008 | 6.07±0.08 | 115±6 | 121±6 | 5.00 | 2.38±0.04 | 4.99±1.17 | *b* |
| 3C98-1 | 179.86 | -31.09 | 0.028±0.003 | 5.40±0.65 | 216±7 | 229±7 | 0.63 | 1.83±0.35 | 1.19±0.32 | *b* |
| 3C98-2 | 179.83 | -31.02 | 0.090±0.004 | 3.16±0.14 | 59±6 | 59±6 | 0.33 | 1.15±0.07 | 0.16±0.05 | *b* |
| 3C98-2 | 179.83 | -31.02 | 0.203±0.013 | 1.39±0.10 | 24±16 | 24±16 | 0.13 | 0.38±0.18 | 0.03±0.03 | *b* |
| 3C98-2 | 179.83 | -31.02 | 0.452±0.011 | 4.62±0.06 | 100±9 | 106±11 | 4.08 | 1.72±0.04 | 3.55±1.00 | *b* |
| 3C98-2 | 179.83 | -31.02 | 0.035±0.003 | 4.65±0.42 | 166±7 | 176±7 | 0.52 | 1.52±0.22 | 0.75±0.19 | *b* |
| 3C98 | 179.84 | -31.05 | 0.092±0.004 | 4.87±0.22 | 17±5 | 17±5 | 0.15 | 2.03±0.10 | 0.02±0.01 | *b* |
| 3C98 | 179.84 | -31.05 | 0.297±0.017 | 1.35±0.09 | 28±20 | 28±20 | 0.21 | 0.32±0.28 | 0.05±0.07 | *b* |
| 3C98 | 179.84 | -31.05 | 0.508±0.012 | 5.21±0.07 | 97±6 | 103±9 | 5.02 | 2.01±0.04 | 4.24±1.10 | *b* |
| 3C98 | 179.84 | -31.05 | 0.041±0.003 | 6.82±0.58 | 134±4 | 142±5 | 0.72 | 2.69±0.27 | 0.84±0.21 | *b* |
| 3C105 | 187.63 | -33.61 | 0.057±0.004 | 4.37±0.44 | 81±6 | 81±6 | 0.40 | 1.66±0.21 | 0.27±0.08 | *b* |
| 3C105 | 187.63 | -33.61 | 2.103±0.054 | 1.81±0.04 | 29±12 | 29±12 | 2.17 | 0.59±0.09 | 0.52±0.44 | *b* |
| 3C105 | 187.63 | -33.61 | 0.169±0.012 | 9.47±0.44 | 160±4 | 169±5 | 4.98 | 3.84±0.20 | 6.93±1.62 | *b* |
| 3C105 | 187.63 | -33.61 | 2.883±0.075 | 1.97±0.04 | 35±10 | 35±10 | 3.87 | 0.64±0.07 | 1.11±0.65 | *b* |
| 3C109 | 181.83 | -27.78 | 0.381±0.040 | 3.48±0.26 | 70±4 | 70±4 | 1.82 | 1.27±0.13 | 1.05±0.29 | *b* |
| 3C109 | 181.83 | -27.78 | 1.979±0.037 | 4.35±0.18 | 73±3 | 73±3 | 12.19 | 1.68±0.08 | 7.30±1.66 | *b* |
| 3C109 | 181.83 | -27.78 | 0.344±0.028 | 2.35±0.16 | 95±3 | 101±5 | 1.50 | 0.40±0.18 | 1.24±0.31 | *b* |
| 3C131 | 171.44 | -7.80 | 0.321±0.006 | 6.95±0.21 | 78±5 | 78±5 | 3.38 | 2.84±0.09 | 2.16±0.55 | *b* |
| 3C131 | 171.44 | -7.80 | 2.152±0.034 | 4.24±0.04 | 44±8 | 44±8 | 7.90 | 1.70±0.03 | 2.89±1.21 | *b* |
| 3C141.0 | 174.53 | -1.31 | 0.260±0.010 | 15.41±0.74 | 255±2 | 271±2 | 19.89 | 6.37±0.32 | 44.31±9.75 | *b* |
| 3C141.0 | 174.53 | -1.31 | 0.875±0.024 | 6.19±0.14 | 36±2 | 36±2 | 3.75 | 2.57±0.06 | 1.10±0.26 | *b* |
| 3C142.1 | 197.62 | -14.51 | 0.101±0.007 | 3.24±0.26 | 9±14 | 9±14 | 0.06 | 1.35±0.12 | 0.005±0.014 | *b* |
| 3C142.1 | 197.62 | -14.51 | 2.362±0.045 | 3.13±0.03 | 49±16 | 49±16 | 7.02 | 1.17±0.06 | 2.81±1.90 | *b* |
| 3C142.1 | 197.62 | -14.51 | 0.203±0.007 | 4.12±0.17 | 23±11 | 23±11 | 0.37 | 1.69±0.08 | 0.07±0.07 | *b* |
| 3C142.1 | 197.62 | -14.51 | 0.083±0.007 | 3.39±0.32 | 119±10 | 126±11 | 0.65 | 1.02±0.20 | 0.67±0.20 | *b* |
| 3C167 | 207.31 | 1.15 | 0.252±0.008 | 21.71±0.98 | 171±7 | 181±7 | 18.27 | 9.14±0.42 | 27.23±6.32 | *b* |
| 3C167 | 207.31 | 1.15 | 0.941±0.024 | 8.01±0.23 | 83±13 | 83±15 | 12.16 | 3.30±0.10 | 8.28±3.32 | *b* |
| 3C167 | 207.31 | 1.15 | 0.386±0.032 | 2.09±0.19 | 27±6 | 27±6 | 0.42 | 0.75±0.10 | 0.09±0.04 | *b* |
| 3C172.0 | 191.20 | 13.41 | 0.059±0.005 | 4.33±0.46 | 59±6 | 59±6 | 0.29 | 1.70±0.21 | 0.14±0.05 | *b* |
| 3C172.0 | 191.20 | 13.41 | 0.031±0.004 | 5.62±1.03 | 33±7 | 33±7 | 0.11 | 2.33±0.45 | 0.03±0.02 | *b* |
| 3C192 | 197.91 | 26.41 | 0.068±0.002 | 4.30±0.12 | 82±4 | 82±4 | 0.47 | 1.63±0.06 | 0.32±0.07 | *b* |
| 3C207 | 212.97 | 30.14 | 0.250±0.002 | 5.25±0.05 | 25±5 | 25±5 | 0.63 | 2.18±0.02 | 0.13±0.06 | *b* |
| 3C207 | 212.97 | 30.14 | 0.298±0.004 | 2.43±0.03 | 20±3 | 20±3 | 0.28 | 0.95±0.02 | 0.04±0.02 | *b* |
| 3C228.0 | 220.8 | 46.6 | 0.082±0.002 | 2.86±0.08 | 70±3 | 70±3 | 0.32 | 0.95±0.05 | 0.19±0.04 | *b* |
| 3C228.0 | 220.8 | 46.6 | 0.027±0.002 | 2.13±0.20 | 44±5 | 44±5 | 0.05 | 0.67±0.12 | 0.02±0.01 | *b* |
| 3C272.1 | 278.21 | 74.48 | 0.029±0.001 | 3.81±0.18 | 99±3 | 104±3 | 0.21 | 1.32±0.09 | 0.18±0.04 | *b* |
| 3C272.1 | 278.21 | 74.48 | 0.009±0.001 | 3.12±0.52 | 112±6 | 119±7 | 0.06 | 0.88±0.33 | 0.06±0.02 | *b* |
| 3C272.1 | 278.21 | 74.48 | 0.015±0.001 | 3.61±0.34 | 87±4 | 87±5 | 0.09 | 1.28±0.18 | 0.06±0.02 | *b* |

| Source (name) | Longitude (deg.) | Latitude (deg.) | $\tau_{peak}(\Delta\tau_{peak})$ | $V_{FHWM}(\Delta V_{FWHM})$ (km sec$^{-1}$) | $T_s(\Delta T_s)$ (K) | $T_k(\Delta T_k)$ (K) | $N_{HI}$ ($\times 10^{20}$ cm$^{-2}$) | $\sigma_{nth}(\Delta\sigma_{nth})$ (km sec$^{-1}$) | $L(\Delta L)$ (pc) | Ref. |
|---|---|---|---|---|---|---|---|---|---|---|
| 3C274.1 | 269.87 | 83.16 | 0.009±0.001 | 3.91±0.51 | 97±9 | 102±12 | 0.07 | 1.38±0.26 | 0.06±0.02 | b |
| 3C274.1 | 269.87 | 83.16 | 0.102±0.001 | 2.94±0.04 | 39±2 | 39±2 | 0.23 | 1.11±0.02 | 0.08±0.02 | b |
| 3C310 | 38.50 | 60.21 | 0.620±0.003 | 1.75±0.01 | 39±3 | 39±3 | 0.82 | 0.48±0.03 | 0.26±0.07 | b |
| 3C310 | 38.50 | 60.21 | 0.061±0.001 | 5.11±0.13 | 48±5 | 48±5 | 0.29 | 2.08±0.06 | 0.12±0.03 | b |
| 3C315 | 39.36 | 58.30 | 0.784±0.011 | 2.15±0.02 | 44±2 | 44±2 | 1.45 | 0.68±0.02 | 0.53±0.12 | b |
| 3C315 | 39.36 | 58.30 | 0.146±0.004 | 4.41±0.15 | 61±15 | 61±15 | 0.77 | 1.73±0.08 | 0.39±0.21 | b |
| 3C318 | 29.64 | 55.42 | 0.482±0.013 | 1.77±0.03 | 34±4 | 34±4 | 0.57 | 0.53±0.04 | 0.16±0.05 | b |
| 3C318 | 29.64 | 55.42 | 0.300±0.011 | 3.36±0.04 | 60±6 | 60±6 | 1.18 | 1.24±0.03 | 0.58±0.17 | b |
| 3C333 | 37.30 | 42.97 | 0.993±0.010 | 2.11±0.02 | 27±8 | 27±8 | 1.10 | 0.76±0.04 | 0.24±0.15 | b |
| 3C348 | 23.05 | 28.95 | 0.259±0.003 | 1.65±0.03 | 12±5 | 11±5 | 0.10 | 0.63±0.03 | 0.01±0.01 | b |
| 3C348 | 23.05 | 28.95 | 0.604±0.004 | 2.12±0.01 | 33±6 | 33±6 | 0.81 | 0.74±0.03 | 0.22±0.09 | b |
| 3C348 | 23.05 | 28.95 | 0.078±0.002 | 3.73±0.09 | 113±2 | 120±2 | 0.64 | 1.24±0.05 | 0.63±0.14 | b |
| 3C353 | 21.20 | 19.64 | 0.006±0.001 | 4.19±0.57 | 292±48 | 311±52 | 0.14 | 0.77±0.62 | 0.36±0.16 | b |
| 3C353 | 21.20 | 19.64 | 1.209±0.007 | 2.80±0.01 | 37±10 | 37±10 | 2.43 | 1.05±0.04 | 0.74±0.43 | b |
| 3C353 | 21.20 | 19.64 | 0.195±0.008 | 5.84±0.07 | 161±5 | 171±5 | 3.59 | 2.18±0.04 | 5.04±1.12 | b |
| 3C353 | 21.20 | 19.64 | 0.859±0.006 | 1.69±0.01 | 27±13 | 27±13 | 0.76 | 0.54±0.10 | 0.17±0.17 | b |
| 3C353 | 21.20 | 19.64 | 0.040±0.001 | 3.01±0.07 | 40±5 | 40±5 | 0.09 | 1.14±0.04 | 0.03±0.01 | b |
| 3C454.0 | 87.35 | -35.65 | 0.045±0.001 | 6.39±0.20 | 156±6 | 165±6 | 0.87 | 2.45±0.09 | 1.19±0.27 | b |
| 3C454.0 | 87.35 | -35.65 | 0.093±0.001 | 3.12±0.05 | 66±2 | 66±2 | 0.38 | 1.10±0.03 | 0.21±0.05 | b |
| 4C07.32 | 320.42 | 69.07 | 0.177±0.005 | 2.19±0.07 | 43±2 | 43±2 | 0.32 | 0.72±0.04 | 0.11±0.03 | b |
| 4C13.65 | 39.31 | 17.72 | 0.022±0.004 | 5.57±1.04 | 43±2 | 43±2 | 0.10 | 2.29±0.46 | 0.04±0.01 | b |
| 4C13.65 | 39.31 | 17.72 | 0.600±0.039 | 2.04±0.08 | 32±3 | 32±3 | 0.75 | 0.70±0.04 | 0.19±0.05 | b |
| 4C13.65 | 39.31 | 17.72 | 0.344±0.023 | 2.57±0.20 | 35±3 | 35±3 | 0.60 | 0.95±0.10 | 0.17±0.05 | b |
| 4C13.67 | 43.50 | 9.15 | 1.161±0.015 | 2.10±0.02 | 26±7 | 26±7 | 1.22 | 0.76±0.04 | 0.26±0.15 | b |
| 4C13.67 | 43.50 | 9.15 | 1.019±0.010 | 4.01±0.04 | 30±8 | 30±8 | 2.36 | 1.63±0.03 | 0.58±0.35 | b |
| 4C13.67 | 43.50 | 9.15 | 0.030±0.003 | 8.20±0.89 | 81±7 | 81±7 | 0.38 | 3.39±0.39 | 0.25±0.08 | b |
| 4C20.33 | 19.54 | 67.46 | 0.058±0.003 | 2.48±0.12 | 52±3 | 52±3 | 0.15 | 0.83±0.07 | 0.06±0.02 | b |
| 4C20.33 | 19.54 | 67.46 | 0.049±0.002 | 4.75±0.21 | 85±3 | 85±3 | 0.38 | 1.84±0.10 | 0.26±0.06 | b |
| 4C22.12 | 188.05 | 0.05 | 1.756±0.045 | 11.63±0.21 | 82±27 | 82±30 | 32.79 | 4.87±0.09 | 22.24±16.11 | b |
| 4C22.12 | 188.05 | 0.05 | 0.840±0.034 | 3.97±0.16 | 34±12 | 34±12 | 2.23 | 1.60±0.08 | 0.63±0.46 | b |
| CTA21 | 166.64 | -33.60 | 0.389±0.006 | 4.18±0.06 | 72±9 | 72±9 | 2.28 | 1.60±0.04 | 1.35±0.44 | b |
| CTA21 | 166.64 | -33.60 | 0.193±0.012 | 1.70±0.09 | 32±12 | 32±12 | 0.20 | 0.51±0.11 | 0.05±0.04 | b |
| CTA21 | 166.64 | -33.60 | 0.108±0.002 | 4.17±0.11 | 74±7 | 74±7 | 0.65 | 1.59±0.06 | 0.39±0.11 | b |
| NRAO140 | 159.00 | -18.76 | 7.011±0.300 | 3.66±0.03 | 27±13 | 27±13 | 13.42 | 1.48±0.04 | 2.96±2.89 | b |
| P0320+05 | 176.98 | -40.84 | 0.478±0.014 | 5.10±0.11 | 56±4 | 56±4 | 2.67 | 2.06±0.05 | 1.23±0.33 | b |
| P0320+05 | 176.98 | -40.84 | 0.223±0.024 | 1.37±0.18 | 38±8 | 38±8 | 0.23 | 0.15±0.37 | 0.07±0.04 | b |
| P0320+05 | 176.98 | -40.84 | 0.372±0.012 | 3.04±0.09 | 53±4 | 53±4 | 1.17 | 1.11±0.05 | 0.51±0.13 | b |
| P0320+05 | 176.98 | -40.84 | 0.408±0.012 | 2.82±0.08 | 44±8 | 44±8 | 0.98 | 1.04±0.05 | 0.35±0.15 | b |
| P0347+05 | 182.27 | -35.73 | 1.051±0.027 | 3.54±0.08 | 62±8 | 62±8 | 4.49 | 1.32±0.05 | 2.29±0.78 | b |
| P0347+05 | 182.27 | -35.73 | 0.171±0.012 | 4.87±0.64 | 172±6 | 182±6 | 2.78 | 1.67±0.34 | 4.15±1.11 | b |
| P0428+20 | 176.81 | -18.56 | 1.132±0.016 | 3.42±0.04 | 29±20 | 29±20 | 2.15 | 1.37±0.06 | 0.50±0.70 | b |
| P0428+20 | 176.81 | -18.56 | 1.318±0.040 | 3.70±0.05 | 33±18 | 33±18 | 3.14 | 1.48±0.06 | 0.85±0.97 | b |
| P0428+20 | 176.81 | -18.56 | 2.346±0.128 | 1.50±0.05 | 23±9 | 23±9 | 1.61 | 0.46±0.09 | 0.31±0.26 | b |

| Source (name) | Longitude (deg.) | Latitude (deg.) | $\tau_{peak}(\Delta\tau_{peak})$ | $V_{FHWM}(\Delta V_{FWHM})$ (km sec$^{-1}$) | $T_s(\Delta T_s)$ (K) | $T_k(\Delta T_k)$ (K) | $N_{HI}$ ($\times 10^{20}$ cm$^{-2}$) | $\sigma_{nth}(\Delta\sigma_{nth})$ (km sec$^{-1}$) | $L(\Delta L)$ (pc) | Ref. |
|---|---|---|---|---|---|---|---|---|---|---|
| P1055+20 | 222.51 | 63.13 | 0.010±0.001 | 7.94±0.30 | 149±6 | 158±7 | 0.23 | 3.17±0.14 | 0.30±0.07 | *b* |
| P1055+20 | 222.51 | 63.13 | 0.073±0.001 | 2.23±0.02 | 40±1 | 40±1 | 0.13 | 0.75±0.01 | 0.04±0.01 | *b* |
| T0526+24 | 181.36 | -5.19 | 1.080±0.037 | 6.22±0.16 | 46±9 | 46±9 | 6.07 | 2.57±0.08 | 2.32±1.01 | *b* |
| T0526+24 | 181.36 | -5.19 | 1.413±0.076 | 7.86±0.20 | 57±10 | 57±10 | 12.37 | 3.27±0.08 | 5.81±2.41 | *b* |
| T0629+10 | 201.53 | 0.51 | 0.152±0.017 | 1.29±0.16 | 28±7 | 28±7 | 0.11 | 0.26±0.18 | 0.03±0.01 | *b* |
| T0629+10 | 201.53 | 0.51 | 1.605±0.035 | 4.37±0.06 | 36±14 | 36±14 | 4.86 | 1.77±0.04 | 1.42±1.17 | *b* |
| T0629+10 | 201.53 | 0.51 | 0.297±0.005 | 28.11±0.46 | 180±9 | 190±9 | 29.21 | 11.87±0.20 | 45.73±10.67 | *b* |
| T0629+10 | 201.53 | 0.51 | 0.354±0.017 | 3.03±0.16 | 47±20 | 47±20 | 0.98 | 1.13±0.10 | 0.38±0.33 | *b* |
| T0629+10 | 201.53 | 0.51 | 0.271±0.016 | 2.81±0.26 | 61±14 | 61±14 | 0.90 | 0.96±0.15 | 0.45±0.23 | *b* |
| T0629+10 | 201.53 | 0.51 | 1.285±0.042 | 1.70±0.05 | 23±10 | 23±10 | 0.96 | 0.58±0.08 | 0.18±0.17 | *b* |
| 3C067 | 146.822 | -30.696 | 0.10±0.002 | 4.81±0.18 | 69±9 | 69±9 | 0.62 | 1.90±0.08 | 0.35±0.12 | *c* |
| 3C067 | 146.822 | -30.696 | 0.41±0.008 | 2.14±0.04 | 43±8 | 43±8 | 0.74 | 0.69±0.05 | 0.26±0.11 | *c* |
| 3C067 | 146.822 | -30.696 | 0.19±0.005 | 5.99±0.13 | 89±15 | 89±18 | 1.94 | 2.39±0.07 | 1.43±0.60 | *c* |
| 3C068.2 | 147.326 | -26.377 | 0.27±0.000 | 2.58±0.07 | 20±9 | 20±9 | 0.27 | 1.02±0.05 | 0.04±0.04 | *c* |
| 3C068.2 | 147.326 | -26.377 | 0.09±0.004 | 2.63±0.24 | 4±8 | 4±8 | 0.02 | 1.10±0.11 | 0.001±0.003 | *c* |
| 3C068.2 | 147.326 | -26.377 | 0.90±0.040 | 2.04±0.05 | 44±15 | 44±15 | 1.57 | 0.62±0.10 | 0.57±0.40 | *c* |
| 3C068.2 | 147.326 | -26.377 | 0.11±0.017 | 3.88±0.62 | 141±15 | 149±5 | 1.15 | 1.22±0.36 | 1.41±0.44 | *c* |
| 3C108 | 171.872 | -20.117 | 0.01±0.002 | 4.73±1.19 | 64±9 | 64±9 | 0.07 | 1.87±0.54 | 0.04±0.02 | *c* |
| 3C108 | 171.872 | -20.117 | 0.47±0.009 | 1.99±0.05 | 58±15 | 58±15 | 1.07 | 0.48±0.13 | 0.51±0.28 | *c* |
| 3C108 | 171.872 | -20.117 | 1.20±0.010 | 2.10±0.02 | 50±5 | 50±5 | 2.44 | 0.62±0.04 | 0.99±0.29 | *c* |
| 4C+25.14 | 171.372 | -17.162 | 0.02±0.006 | 16.30±5.51 | 13±9 | 13±9 | 0.09 | 6.91±2.34 | 0.01±0.01 | *c* |
| 4C+25.14 | 171.372 | -17.162 | 0.04±0.014 | 4.59±0.88 | 22±8 | 22±8 | 0.09 | 1.90±0.38 | 0.02±0.01 | *c* |
| 4C+25.14 | 171.372 | -17.162 | 0.11±0.011 | 1.82±0.18 | 42±15 | 42±15 | 0.17 | 0.50±0.16 | 0.06±0.04 | *c* |
| 4C+25.14 | 171.372 | -17.162 | 0.14±0.016 | 7.15±0.38 | 131±5 | 138±5 | 2.61 | 2.84±0.17 | 2.96±0.76 | *c* |
| 4C+25.14 | 171.372 | -17.162 | 1.03±0.012 | 2.10±0.03 | 42±7 | 42±7 | 1.77 | 0.67±0.04 | 0.61±0.23 | *c* |
| 4C+25.14 | 171.372 | -17.162 | 0.06±0.004 | 1.76±0.13 | 16±6 | 16±6 | 0.03 | 0.65±0.07 | 0.004±0.003 | *c* |
| 4C+25.14 | 171.372 | -17.162 | 0.03±0.003 | 3.82±0.41 | 32±6 | 32±6 | 0.07 | 1.54±0.18 | 0.02±0.01 | *c* |
| 4C+26.12 | 165.818 | -21.059 | 0.16±0.003 | 2.60±0.05 | 53±9 | 53±9 | 0.44 | 0.88±0.05 | 0.20±0.08 | *c* |
| 4C+26.12 | 165.818 | -21.059 | 0.30±0.005 | 1.83±0.04 | 53±8 | 53±8 | 0.56 | 0.41±0.09 | 0.24±0.09 | *c* |
| 4C+27.07 | 145.012 | -31.093 | 0.10±0.003 | 3.67±0.16 | 88±9 | 88±12 | 0.60 | 1.31±0.09 | 0.43±0.14 | *c* |
| 4C+27.07 | 145.012 | -31.093 | 0.27±0.004 | 2.96±0.05 | 61±8 | 61±8 | 0.94 | 1.04±0.04 | 0.47±0.16 | *c* |
| 4C+27.14 | 175.828 | -9.364 | 0.08±0.007 | 1.74±0.18 | 4±9 | 4±9 | 0.01 | 0.72±0.09 | 0.0003±0.001 | *c* |
| 4C+27.14 | 175.828 | -9.364 | 0.29±0.012 | 2.49±0.11 | 34±8 | 34±8 | 0.47 | 0.91±0.07 | 0.13±0.07 | *c* |
| 4C+27.14 | 175.828 | -9.364 | 0.14±0.005 | 7.65±0.43 | 66±5 | 66±5 | 1.33 | 3.16±0.19 | 0.72±0.19 | *c* |
| 4C+27.14 | 175.828 | -9.364 | 0.23±0.011 | 1.19±0.07 | 14±15 | 14±15 | 0.07 | 0.37±0.17 | 0.008±0.017 | *c* |
| 4C+27.14 | 175.828 | -9.364 | 0.05±0.004 | 3.89±0.42 | 85±7 | 85±7 | 0.32 | 1.42±0.21 | 0.22±0.07 | *c* |
| 4C+27.14 | 175.828 | -9.364 | 0.14±0.016 | 2.84±0.31 | 45±6 | 45±6 | 0.35 | 1.04±0.15 | 0.13±0.05 | *c* |
| 4C+27.14 | 175.828 | -9.364 | 0.21±0.020 | 12.26±0.46 | 162±36 | 171±38 | 8.06 | 5.07±0.20 | 11.34±5.73 | *c* |
| 4C+27.14 | 175.828 | -9.364 | 0.10±0.019 | 2.86±0.74 | 14±6 | 14±6 | 0.07 | 1.17±0.33 | 0.008±0.008 | *c* |
| 4C+27.14 | 175.828 | -9.364 | 1.10±0.022 | 2.18±0.05 | 13±18 | 13±18 | 0.60 | 0.87±0.09 | 0.06±0.18 | *c* |
| 4C+28.06 | 148.781 | -28.443 | 0.28±0.004 | 5.31±0.07 | 70±8 | 70±8 | 2.02 | 2.12±0.04 | 1.16±0.37 | *c* |
| 4C+28.06 | 148.781 | -28.443 | 0.24±0.007 | 1.85±0.06 | 27±15 | 27±15 | 0.24 | 0.63±0.10 | 0.05±0.06 | *c* |
| 4C+28.06 | 148.781 | -28.443 | 0.03±0.003 | 1.70±0.23 | 37±5 | 37±5 | 0.04 | 0.46±0.16 | 0.01±0.00 | *c* |

| Source (name) | $l$ (deg.) | $b$ (deg.) | $\tau_{peak}(\Delta\tau_{peak})$ | $V_{FWHM}(\Delta V_{FWHM})$ (km sec$^{-1}$) | $T_s(\Delta T_s)$ (K) | $T_k(\Delta T_k)$ (K) | $N_{HI}$ ($\times 10^{20}$ cm$^{-2}$) | $\sigma_{nth}(\Delta\sigma_{nth})$ (km sec$^{-1}$) | $L(\Delta L)$ (pc) | Ref. |
|---|---|---|---|---|---|---|---|---|---|---|
| 4C+28.07 | 149.466 | -28.528 | 0.15±0.009 | 2.36±0.10 | 58±8 | 58±8 | 0.40 | 0.72±0.08 | 0.19±0.07 | c |
| 4C+28.07 | 149.466 | -28.528 | 0.57±0.019 | 3.02±0.08 | 16±15 | 16±15 | 0.55 | 1.23±0.06 | 0.07±0.14 | c |
| 4C+28.07 | 149.466 | -28.528 | 0.19±0.026 | 2.94±0.25 | 39±5 | 39±5 | 0.42 | 1.11±0.12 | 0.14±0.05 | c |
| 4C+28.07 | 149.466 | -28.528 | 0.02±0.003 | 6.25±1.93 | 139±7 | 148±7 | 0.28 | 2.41±0.90 | 0.34±0.14 | c |
| 4C+29.05 | 140.716 | -30.876 | 0.03±0.003 | 2.20±0.31 | 33±9 | 33±9 | 0.04 | 0.78±0.17 | 0.01±0.01 | c |
| 4C+29.05 | 140.716 | -30.876 | 0.14±0.003 | 3.93±0.08 | 49±8 | 49±8 | 0.54 | 1.54±0.04 | 0.22±0.09 | c |
| 4C+33.10 | 169.055 | -7.572 | 0.12±0.006 | 2.36±0.13 | 34±9 | 34±9 | 0.19 | 0.85±0.08 | 0.05±0.03 | c |
| 4C+33.10 | 169.055 | -7.572 | 0.20±0.006 | 3.06±0.09 | 52±8 | 52±8 | 0.62 | 1.12±0.05 | 0.27±0.10 | c |
| 4C+33.10 | 169.055 | -7.572 | 0.07±0.005 | 2.77±0.25 | 45±15 | 45±15 | 0.17 | 1.00±0.14 | 0.06±0.04 | c |
| 4C+33.10 | 169.055 | -7.572 | 0.21±0.005 | 4.27±0.17 | 19±5 | 19±5 | 0.33 | 1.77±0.07 | 0.05±0.03 | c |
| 4C+33.10 | 169.055 | -7.572 | 0.62±0.010 | 3.57±0.12 | 48±7 | 48±7 | 2.07 | 1.38±0.06 | 0.82±0.29 | c |
| 4C+33.10 | 169.055 | -7.572 | 0.95±0.012 | 4.21±0.21 | 70±6 | 70±6 | 5.43 | 1.62±0.10 | 3.11±0.88 | c |
| 4C+33.10 | 169.055 | -7.572 | 0.83±0.028 | 3.03±0.11 | 70±6 | 70±6 | 3.44 | 1.04±0.06 | 1.98±0.55 | c |
| 4C+33.10 | 169.055 | -7.572 | 0.19±0.007 | 2.46±0.14 | 37±36 | 37±36 | 0.34 | 0.89±0.18 | 0.10±0.20 | c |
| 4C+34.07 | 144.312 | -24.550 | 0.24±0.007 | 1.84±0.05 | 21±9 | 21±9 | 0.18 | 0.66±0.06 | 0.03±0.03 | c |
| 4C+34.07 | 144.312 | -24.550 | 0.09±0.006 | 1.79±0.11 | 64±15 | 64±15 | 0.21 | 0.23±0.31 | 0.11±0.06 | c |
| 4C+34.09 | 150.936 | -20.486 | 0.22±0.004 | 1.69±0.03 | 5±9 | 5±9 | 0.03 | 0.69±0.05 | 0.001±0.004 | c |
| 4C+34.09 | 150.936 | -20.486 | 0.19±0.002 | 9.80±0.11 | 157±15 | 166±16 | 5.79 | 3.99±0.05 | 7.92±2.24 | c |
| 5C06.237 | 143.882 | -26.525 | 0.40±0.004 | 3.42±0.07 | 52±9 | 52±9 | 1.40 | 1.30±0.04 | 0.60±0.24 | c |
| 5C06.237 | 143.882 | -26.525 | 0.10±0.006 | 2.18±0.16 | 62±8 | 62±8 | 0.27 | 0.59±0.12 | 0.14±0.05 | c |
| B20218+35 | 142.602 | -23.487 | 0.03±0.011 | 2.77±0.50 | 116±9 | 123±9 | 0.19 | 0.61±0.42 | 0.19±0.09 | c |
| B20218+35 | 142.602 | -23.487 | 0.04±0.030 | 6.24±2.03 | 50±15 | 50±15 | 0.26 | 2.57±0.89 | 0.11±0.11 | c |
| B20218+35 | 142.602 | -23.487 | 0.11±0.031 | 2.82±0.34 | 62±8 | 62±8 | 0.38 | 0.96±0.18 | 0.19±0.09 | c |
| B20400+25 | 168.026 | -19.648 | 0.31±0.007 | 2.17±0.04 | 67±15 | 67±15 | 0.88 | 0.54±0.12 | 0.49±0.24 | c |
| B20400+25 | 168.026 | -19.648 | 0.16±0.006 | 4.79±0.10 | 54±9 | 54±9 | 0.80 | 1.92±0.05 | 0.36±0.14 | c |
| B20411+34 | 163.796 | -11.982 | 0.02±0.004 | 3.86±0.66 | 15±9 | 15±9 | 0.03 | 1.60±0.29 | 0.004±0.005 | c |
| B20411+34 | 163.796 | -11.982 | 0.07±0.003 | 5.27±0.35 | 106±8 | 112±9 | 0.74 | 2.02±0.17 | 0.68±0.19 | c |
| B20411+34 | 163.796 | -11.982 | 0.90±0.128 | 4.73±0.20 | 39±15 | 39±15 | 3.27 | 1.93±0.09 | 1.05±0.84 | c |
| B20411+34 | 163.796 | -11.982 | 0.18±0.073 | 6.18±1.27 | 69±5 | 69±5 | 1.53 | 2.51±0.56 | 0.87±0.45 | c |
| NV0157+28 | 139.899 | -31.835 | 0.05±0.001 | 9.76±0.18 | 302±9 | 321±10 | 2.90 | 3.81±0.08 | 7.67±1.69 | c |
| NV0157+28 | 139.899 | -31.835 | 0.04±0.002 | 1.96±0.12 | 26±8 | 26±8 | 0.04 | 0.69±0.08 | 0.01±0.01 | c |
| NV0232+34 | 145.598 | -23.984 | 0.13±0.010 | 2.04±0.16 | 43±8 | 43±8 | 0.22 | 0.63±0.11 | 0.08±0.04 | c |
| NV0232+34 | 145.598 | -23.984 | 0.06±0.014 | 8.49±0.71 | 321±9 | 342±10 | 3.10 | 3.19±0.34 | 8.71±2.88 | c |
| NV0232+34 | 145.598 | -23.984 | 0.20±0.012 | 1.82±0.09 | 37±15 | 37±15 | 0.27 | 0.54±0.13 | 0.08±0.07 | c |
| NV0232+34 | 145.598 | -23.984 | 0.09±0.008 | 2.11±0.19 | 31±5 | 31±5 | 0.11 | 0.74±0.10 | 0.03±0.01 | c |
| B0438-436 | 248.411 | -41.565 | 0.00108±0.000 | 15.0±2.9 | 1000±345 | 1082±417 | 0.30 | 5.62±1.43 | 2.67±2.09 | d |